\documentclass[12pt]{article}
\usepackage{dcolumn} 
\usepackage{epsfig}
\usepackage{cite}
\usepackage{amsmath}
\usepackage{hhline}
\usepackage{amssymb}
\usepackage{times}
\usepackage{rotating}

\newlength{\dinwidth}
\newlength{\dinmargin}
\def\Journal#1#2#3#4{{#1} {\bf #2} (#3) #4}

\def\ar#1#2#3   {\Journal{\em Ann. Rev. Nucl. Part. Sci.}{\bf#1}{#2}{#3}}
\def\err#1#2#3  {\Journal{\em Erratum}{\bf#1}{#2}{#3}}
\def\ib#1#2#3   {\Journal{ibid.}{\bf#1}{#2}{#3}}
\def\ijmp#1#2#3 {\Journal{\em Int. J. Mod. Phys.}{\bf#1}{#2}{#3}}
\def\jetp#1#2#3 {\Journal{\em JETP Lett.}{\bf#1}{#2}{#3}}
\def\mpl#1#2#3  {\Journal{\em Mod. Phys. Lett.}{\bf#1}{#2}{#3}}
\def\nim#1#2#3  {\Journal{\em Nucl. Instrum. Meth.}{\bf#1}{#2}{#3}}
\def\nc#1#2#3   {\Journal{\em Nuovo Cim.}{\bf#1}{#2}{#3}}
\def\np#1#2#3   {\Journal{\em Nucl. Phys.}{\bf#1}{#2}{#3}}
\def\pl#1#2#3   {\Journal{\em Phys. Lett.}{\bf#1}{#2}{#3}}
\def\prep#1#2#3 {\Journal{\em Phys. Rep.}{\bf#1}{#2}{#3}}
\def\prev#1#2#3 {\Journal{\em Phys. Rev.}{\bf#1}{#2}{#3}}
\def\prl#1#2#3  {\Journal{\em Phys. Rev. Lett.}{\bf#1}{#2}{#3}}
\def\ptp#1#2#3  {\Journal{\em Prog. Th. Phys.}{\bf#1}{#2}{#3}}
\def\rmp#1#2#3  {\Journal{\em Rev. Mod. Phys.}{\bf#1}{#2}{#3}}
\def\rpp#1#2#3  {\Journal{\em Rep. Prog. Phys.}{\bf#1}{#2}{#3}}
\def\sjnp#1#2#3 {\Journal{\em Sov. J. Nucl. Phys.}{\bf#1}{#2}{#3}}
\def\spj#1#2#3  {\Journal{\em Sov. Phys. JEPT}{\bf#1}{#2}{#3}}
\def\zp#1#2#3   {\Journal{\em Z. Phys.}{\bf#1}{#2}{#3}}
\def\ejc#1#2#3  {\Journal{\em Eur. Phys. J.}{\bf#1}{#2}{#3}}
\def\jetp2#1#2#3 {\Journal{\em J. Exp. Theor. Phys.}{\bf#1}{#2}{#3}}
\def\cpc#1#2#3 {\Journal{\em Comput. Phys. Commun.}{\bf#1}{#2}{#3}}

\def\gsim{\ \,\lower.25ex\hbox{$\scriptstyle\sim$}\kern-1.30ex%
\raise 0.55ex\hbox{$\scriptstyle >$}\ \,}
\def\lsim{\ \,\lower.25ex\hbox{$\scriptstyle\sim$}\kern-1.30ex%
\raise 0.55ex\hbox{$\scriptstyle <$}\ \,}

\newcommand{\ptst}{$p_{t,\psi}^{*2}$}
\newcommand{\ptstw}{p_{t,\psi}^{*2}}
\newcommand{\ptstar}{\ensuremath{p_{t,\psi}^*}}
\newcommand{\ptt}{\ensuremath{p_{t,\psi}^2} }

\newcommand{\bdec}{$b\ra\jpsiw+X$}
\newcommand{\ccbar}{\ensuremath{\mit c\overline{c}}}
\newcommand{\cms}{centre of mass}

\newcommand{\dsdq}{$d\sigma/d\qsq$}
\newcommand{\dsdpts}{$d\sigma/d\ptstw$}
\newcommand{\ppbar}{$p\overline{p}$}
\newcommand{\bbbar}{b\overline{b}}

\newcommand{\ra}{\rightarrow}
\newcommand{\picb}{\mbox{pb}^{-1}}
\newcommand{\gev}{\,\mbox{\rm GeV}}
\newcommand{\GeV}{\,\mbox{\rm GeV}}

\newcommand{\gevt}{\,\mbox{\GeV$^2$}}

\newcommand{\qsq}{\ensuremath{Q^2}}

\newcommand{\jpsi}{$J/\psi$}
\newcommand{\jpsiw}{J/\psi}

\newcommand{\gstp}{$\gamma^* p$}

\newcommand{\csm}{Colour Singlet Model}
\newcommand{\colsing}{colour singlet}
\newcommand{\coloct}{colour octet}
\newcommand{\psits}{\ensuremath{\psi(2S)}}
\begin{document}

\begin{titlepage}

%\begin{flushleft}
\noindent
%Date:   \today; \the\time            \\
%Version:  3.5        \\
%Editors:  S. Mohrdieck, B. Naroska          \\
%Referees:  C. Grab, F. Sefkow         \\
DESY 02-060  \hfill  ISSN 0418-9833 \\
May 2002

\vspace*{3cm}

\begin{center}
\begin{Large}

{\bf \boldmath Inelastic Leptoproduction of $J/\psi$ Mesons at HERA}
%\\ in the range $2<\qsq<100\,\gevt$}

\vspace{2cm}

H1 Collaboration

\end{Large}
\end{center}

\vspace{2cm}

\begin {abstract}
\noindent
The leptoproduction of $J/\psi$ mesons is studied in inelastic reactions 
for four momentum transfers $2<Q^2<100\mbox{~GeV}^2$. The data
were taken with the H1 detector at the electron proton collider HERA and 
correspond to an integrated luminosity of $77\,\picb$. 
Single differential and double differential cross sections are measured with increased 
precision compared with previous analyses. New
leading order calculations within 
the non-relativistic QCD factorisation approach including colour octet and
colour singlet contributions are compared with the data and are found to give a 
reasonable description of most distributions. An exception is
the shape of the distribution in the $J/\psi$ fractional energy, $z$, 
which deviates significantly from that of the data.
Comparisons with photoproduction are made and the polarisation of the
produced \jpsi\ meson is analysed.
\end{abstract}

\vspace{1.5cm}

\begin{center}
To be submitted to Eur.~Phys.~J.~C
\end{center}
\vspace{1.5cm}
\end{titlepage}
%---------------------------------------------
\begin{flushleft}
  %-- H1AUTS Author list by names 
%-- Status: Thu Nov  1 16:52:07 MET 2001  Number of authors = 329 

C.~Adloff$^{33}$,              %WUPP-LEFT      07/01           Adloff              
V.~Andreev$^{24}$,             %LPI -PD        8/88            Andreev             
B.~Andrieu$^{27}$,             %ECPL-LEFT      09/01           Andrieu             
T.~Anthonis$^{4}$,             %ANTW-ST        11/99           Anthonis            
A.~Astvatsatourov$^{35}$,      %ZEUT-ST        02/98           Astvatsatourov      
A.~Babaev$^{23}$,              %ITEP-PD        8/88            Babaev              
J.~B\"ahr$^{35}$,              %ZEUT-PD        8/88            Baehr               
P.~Baranov$^{24}$,             %LPI -PD        8/88            Baranovp            
E.~Barrelet$^{28}$,            %PARI-PD        11/99           Barrelet            
W.~Bartel$^{10}$
% ...

%-- H1 Institutes 
\bigskip{\it
% ...
$ ^{4}$ Inter-University Institute for High Energies ULB-VUB, Brussels;
          Universitaire Instelling Antwerpen, Wilrijk; Belgium$^{ c}$ \\
% ...
$ ^{10}$ DESY, Hamburg, Germany \\
% ...
$ ^{23}$ Institute for Theoretical and Experimental Physics,
          Moscow, Russia$^{ l}$ \\
 $ ^{24}$ Lebedev Physical Institute, Moscow, Russia$^{ e,h}$ \\
% ...
$ ^{27}$ LPNHE, Ecole Polytechnique, IN2P3-CNRS, Palaiseau, France \\
 $ ^{28}$ LPNHE, Universit\'{e}s Paris VI and VII, IN2P3-CNRS,
          Paris, France \\
% ...
$ ^{33}$ Fachbereich Physik, Bergische Universit\"at Gesamthochschule
          Wuppertal, Wuppertal, Germany \\
% ...
$ ^{35}$ DESY, Zeuthen, Germany \\
% ...
$ ^c$ Supported by FNRS-NFWO, IISN-IIKW \\
% ...
$ ^e$ Supported by the Deutsche Forschungsgemeinschaft \\
% ...
$ ^h$ Supported by Russian Foundation for Basic Research
      grant no. 96-02-00019 \\
% ...
$ ^l$ Partially Supported by Russian Foundation
      for Basic Research, grant    no. 00-15-96584 \\
}
\end{flushleft}

\newpage

%-----------------------------------------------------------------------------
\section{Introduction}

\begin{figure}[htbp] 
\begin{center}
\begin{picture}(15.5,8.7)
 \put(2.3,-0.7){\epsfig{file=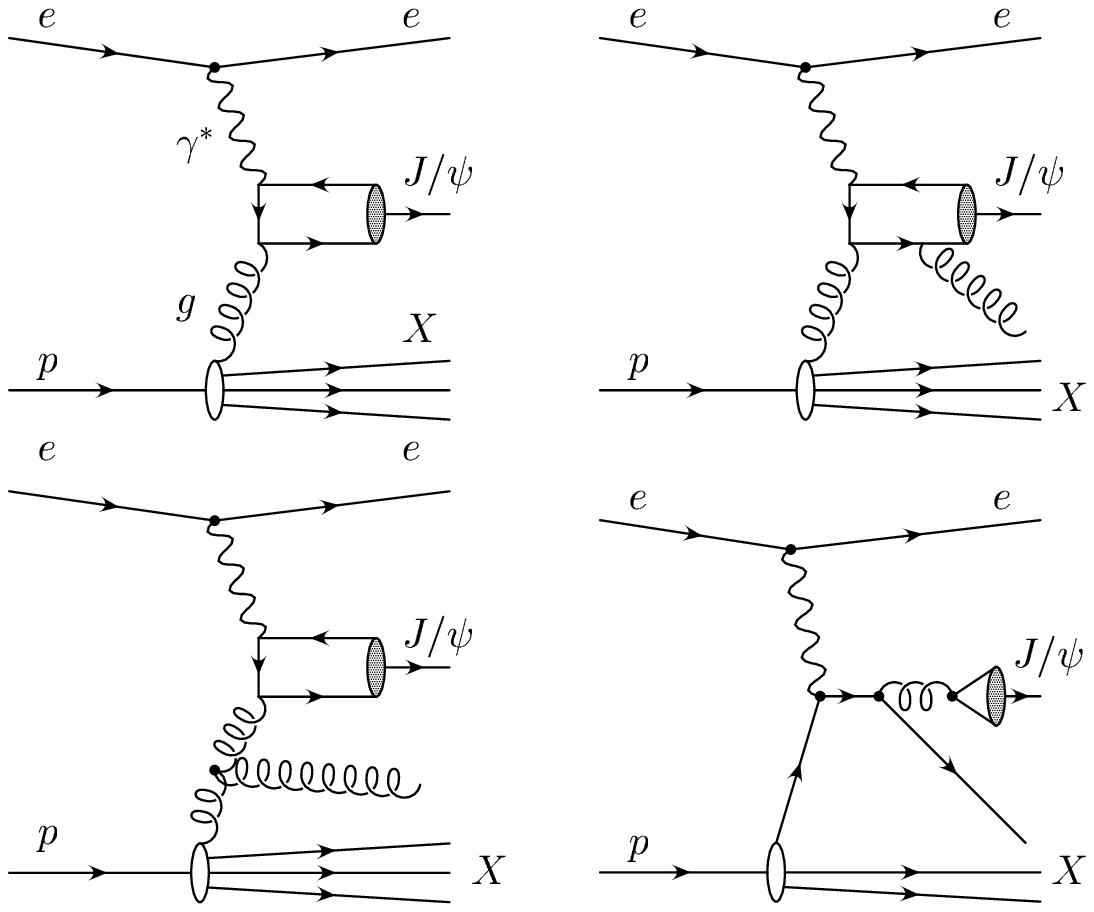,width=11cm,clip=}}
  \put(4.3,8.3){\sf a)}
  \put(10.2,8.3){\sf b)}
 % \put(11.0,7.8){\sf c)}
  \put(4.3,3.7){\sf c)}
  \put(10.2,3.6){\sf d)}
\end{picture}
\end{center}
\caption{Generic diagrams for charmonium production mechanisms: a) Photon gluon fusion via 
a ``$2\ra1$'' process; b--d) ``$2\ra2$'' processes. a--d) contribute via \coloct\ mechanisms, while b) can also contribute in \colsing\ processes.
Additional soft gluons emitted during the hadronisation process are not shown.} 
\label{diag}
\end{figure}

Inelastic leptoproduction of $J/\psi$ mesons at HERA, $e+p\ra e+\jpsiw+X$,
 is dominated by boson gluon fusion, $\gamma^*g\ra\ccbar$. The aim of 
current experimental and theoretical efforts is a detailed understanding of 
this production process.
Before HERA started operation, the limited amount of lepto- and 
photoproduction data~(\cite{Allasia:1991zx} and references therein) was found 
to be 
described by the Colour Singlet Model (CSM)~\cite{colsing}. In the CSM 
the $c\overline{c}$ pair is produced in the hard $\gamma^*g$ interaction 
in the quantum state of the $J/\psi$ meson, i.e. in a colour 
singlet state with spin 1 and no orbital angular momentum. This is possible 
due to the emission of an additional 
hard gluon (see Fig.~\ref{diag}b).   
The process was advocated as a means of determining the gluon density
 in the proton, since it
is calculable in perturbative Quantum Chromodynamics 
(pQCD) using e.g. potential models for the formation of the \jpsi\ meson. 

In recent years the interest in inelastic \jpsi\ production has shifted
considerably since the CSM fails to reproduce the  
production rate of \jpsi\ and \psits\ mesons in \ppbar\ collisions
by more than an order of magnitude \cite{cdf}. 
Nowadays, one of the main aims is the 
investigation of the r\^{o}le of ``colour octet'' contributions, which have 
been invoked to describe the \ppbar\ data.
 Colour octet contributions arise naturally  
in the theoretical description of quarkonium production 
based on non--relativistic QCD and factorisation 
(NRQCD)~\cite{nrqcd}. NRQCD is an effective field theory 
in which the \jpsi\ production process factorises into 
terms for the short distance transition 
(e.g. $\gamma^* g\rightarrow c\overline{c}(g)$) 
and long distance matrix elements (LDMEs) for the transition of the 
$c\overline{c}$ pair into an observable meson.
The \ccbar\ pairs can now be in many different angular momentum states, in 
colour singlet and also in colour octet states, in which case the
transition to the \jpsi\ meson is thought to proceed via soft gluon emission. 
The short distance coefficients are calculable in pQCD and 
a double expansion in the strong coupling parameter $\alpha_s$ and $v$,  
the relative velocity of the quark and antiquark, is obtained.  
Many contributions are possible (examples are shown in Fig.~\ref{diag}) 
and only the most important contributions 
are kept in a specific calculation \cite{Fle972,kniehl}. 
 The leading term in the velocity expansion is the \colsing\ term, so if it is  
assumed that all other terms do not contribute, the CSM is recovered. 
Although the octet LDMEs are  at present not calculable, they are assumed to be  
universal. They have been extracted from the measurement of \jpsi\
production in \ppbar\ collisions by fitting the leading order (LO) theoretical 
calculation to the data (see e.g. \cite{Braaten:2000qk, Kraemer} and
references therein) and are then used in predictions for electroproduction.

First attempts to establish the relative importance of colour octet contributions in 
lepton proton interactions were made in 
the photoproduction limit, $Q^2 \rightarrow 0$ \cite{H1962,Zeus972},
where \qsq\ is the negative squared four momentum transfer. The 
predicted  large contributions at high values of the \jpsi\ fractional
energy, $z$, were not observed. Here, $z$ denotes 
the \jpsi\ energy relative to the photon energy in the proton rest system.
In the analysis of data at high $Q^2$ the dependence of the cross section on 
$Q^2$ may give additional insight into the production process\cite{Fle972}.

Analysing leptoproduction at finite \qsq\ has experimental and theoretical 
advantages compared with photoproduction. At high $Q^2$ theoretical
uncertainties in the models decrease and resolved photon processes are expected
to be negligible. Furthermore background from diffractive production of 
charmonia is expected to decrease faster with \qsq\ 
than the inelastic process.  The distinct signature of the scattered lepton 
makes the process easier to detect.
A first comparison between data and NRQCD calculations in the range 
$2<\qsq<80\,\gevt$ and $40<W<180\gev$ was presented
in~\cite{Adloff:1999zs}, $W$ being the mass of the hadronic
final state or equivalently the centre of mass energy of the photon proton system. The 
NRQCD calculations compared with the data in~\cite{Adloff:1999zs} were performed taking 
into account only ``$2\ra1$'' 
diagrams \cite{Fle972} (compare Fig.~\ref{diag}a) and disagreement between data 
and theory was observed both in the absolute values of the cross sections and
in their shapes as functions of the variables studied. 

In this paper, an analysis of $e+p\ra e+\jpsiw+X$ is presented 
in the kinematic region $2<Q^2<100\mbox{~GeV}^2$ and
$50<W<225\,\mbox{~GeV}$ with increased  
statistics compared to our previous publication~\cite{Adloff:1999zs}.
Differential cross sections are measured for the whole \qsq\ range and
for a subset with $Q^2>12\, \mbox{~GeV}^2$.
The data are compared with theoretical predictions  \cite{kniehl} in the
NRQCD framework taking into account \coloct\ (CO) and
\colsing\ (CS) contributions. In contrast to the previous
NRQCD calculation, diagrams of the type ``$2\ra 2$'' are
taken into account (e.g. diagrams \ref{diag}b, c and d).
The \jpsi\  polarisation is measured by analysing the
decay angular distribution and its \qsq\  
dependence is investigated. The polarisation measurements are compared with 
the prediction of a calculation  \cite{baranov} within 
a ``$k_t$ factorisation'' approach, i.e. allowing transverse momentum 
(``$k_t$'') for the incoming gluon, using unintegrated parton density 
functions and off-shell matrix elements including 
\coloct\ and \colsing\ contributions.

%%%%%%%%%%%%%%%%%%%%%%%%%%%%%%%%%%%%%%%%%

\section{Detector, Kinematics and Simulations} \label{detector}

The data presented were collected in the years 1997--2000 and 
correspond to a total integrated luminosity of $77.0\pm
1.2 \mbox{~pb}^{-1}$. HERA was operated for most of this time 
with $27.5\mbox{~GeV}$ positrons. Roughly 12\% of the data were
taken with electrons of the same energy.
%\footnote{In the following the beam lepton will
%generically be called an electron.}. 
 In 1997 the proton energy was
$820\mbox{~GeV}$. It was increased to $920\mbox{~GeV}$ thereafter (sample of
$\sim63\mbox{~pb}^{-1}$).  

The experimental methods are similar to those described 
in~\cite{Adloff:1999zs} and further details can be found 
in~\cite{Mohrdieck:2000mk}. 
$J/\psi$ mesons are detected via the decays $J/\psi \rightarrow \mu^+\mu^-$ and
$J/\psi \rightarrow e^+e^-$ (branching fractions of $5.88\pm 0.10\%$ and
$5.93 \pm 0.10\%$, respectively \cite{Groom:2000in}).

\subsection{Detector} 
A detailed description of the H1 detector can be found elsewhere
\cite{h1det}. Here we give an overview of the most important components 
for the present analysis.
The central tracking detector (CTD) of H1 consists mainly of two coaxial 
cylindrical drift 
chambers for the measurement of charged particles and their momenta transverse
to the beam direction and  
two polygonal drift chambers for measurement of the $z$ 
coordinates\footnote{H1 uses a right handed coordinate system, the forward 
($+z$) direction, with respect to which the polar angle $\theta$ is 
measured, is defined as that of the proton beam. The 
backward direction ($-z$) is that of the lepton beam.}.
The CTD is situated inside the solenoidal magnet,
which generates a field of $1.15$~T.
The tracking system is complemented in the
forward direction by a set of drift chambers with 
wires perpendicular to the beam direction which allow particle
detection for polar angles $\theta\gsim 7^\circ$. Multiwire proportional
chambers are used for triggering purposes.

In the $Q^2$ range studied here, the scattered lepton is identified through its 
energy deposition in the backward
electromagnetic calorimeter SpaCal \cite{spacal}. The SpaCal signal is 
also used to trigger the events, in conjunction with signals from the tracking chambers. 
A drift chamber (BDC) in front of the SpaCal is used in combination
with the interaction vertex  to reconstruct
the polar angle  $\theta_e$ of the scattered lepton. 
%In the case that no BDC track is found, $\theta_e$ is determined from the
%SpaCal cluster position only.

The liquid argon (LAr) calorimeter surrounds the CTD and is segmented into 
electromagnetic and  hadronic sections. It covers the polar
angular range $4^\circ < \theta < 154^\circ$ with full azimuthal coverage.
The detector is surrounded by an instrumented iron return yoke that is 
used for muon identification (central muon
detector CMD, $4^\circ < \theta < 171^\circ$). 

The \jpsi\ decay electrons are identified via
their energy deposition in the electromagnetic part of the calorimeter
and via their specific 
energy loss in the gas of the central drift chambers.  
Muons are identified as minimum ionising particles in the LAr calorimeter
or through track segments reconstructed in the CMD.

\subsection{Kinematics} \label{kinematics}

The kinematics for charmonium production are described with the standard
variables used for deep inelastic interactions,
namely the square of the $ep$ centre of mass energy, $s = (p+k)^2$,
the squared four momentum transfer $Q^2 = -q^2$ and the mass of the hadronic 
final state $W = \sqrt{(p+q)^2}$. Here $k$, $p$ and
$q$ are the four-momenta of the incident lepton, proton and
virtual photon, respectively. In addition, the scaled energy transfer 
$y=p\cdot q / p \cdot k $ (energy fraction transferred from the lepton to the
hadronic final state in the proton rest frame) and  the  \jpsi\
fractional energy $z = (p_\psi\cdot p)/(q\cdot p)$ are used, where $p_\psi$
denotes the $J/\psi$ four-momentum. 

The event kinematics are reconstructed using a method
which combines the measurement of the scattered lepton
and the hadronic final state to obtain good resolution in the
entire kinematic range. The variable $Q^2= 4\,E\,E'\cos^2 \frac{\theta_e}{2}$
is reconstructed from the energy $E'$ and angle $\theta_e$ of the
scattered lepton ($E$ is the energy of the incoming lepton). 
For the calculation of $y$ and 
$z$ the hadronic final state is used in addition.  Thus

\begin{equation}
  y = \frac{\sum_{had}(E-p_z)}{\sum (E-p_z)} \qquad\mbox{~~~and~~~}\qquad
  z = \frac{p_\psi\cdot p}{y\,s/2}=\frac{(E-p_z)_{\psi}}{\sum_{had}(E-p_z)}, 
\label{yzequation}
\end{equation}
where $\sum (E-p_z)$ runs over all the final state particles including
the scattered lepton, and in
$\sum_{had}(E-p_z)$ only the final state hadrons are summed. The
\jpsi\ momentum is reconstructed from the momenta of the decay
leptons. For the calculation of the sums in equations (\ref{yzequation}) a
combination of tracks reconstructed in the CTD and energy depositions  
in the LAr and SpaCal calorimeters is used.  $W$ is reconstructed
using the relation $W^2 = ys - Q^2$.

%\subsection{Resolution}
Differential cross sections are measured as functions of the
following variables: $Q^2$, $W$, $z$, 
the transverse momentum squared of the \jpsi\ with respect to the beam axis 
$p_{t,\psi}^2$ and
the rapidity \footnote{The rapidity $Y= \frac{1}{2}\ln\frac{E+p_z}{E-p_z}$ 
of the \jpsi\ is calculated with respect to the proton  direction in the laboratory 
frame and with respect to the photon direction in the  photon-proton \cms\ frame.} 
of the $J/\psi$ in the laboratory frame  $Y_{lab}$. Differential cross
sections are also given for $p_{t,\psi}^{*2}$ and $Y^{*}$, which 
are computed in the $\gamma^*p$ \cms\ frame.
The resolution, as determined from the detector simulation, is
 $2-5\%$ for the variables $Q^2$,  $p_{t,\psi}^2$, $Y_{lab}$  and $Y^\ast$. 
For $z$ the resolution is $\sim8\%$ at high $z\sim 1$ degrading to 15\% at 
low $z$ values. For $W$ the resolution is $\sim7\%$ 
for $W<145\gev$ and $\sim4\%$ above. The resolution of $p_{t,\psi}^{*2}$
is somewhat worse ($\sim 30\%$ of the chosen bin widths). 

\subsection{Monte Carlo Simulations} \label{simulation}

Corrections for detector effects are applied to the
data using a Monte Carlo simulation in which the H1 detector response is
simulated in detail. The simulated events are passed through the
same reconstruction and analysis chain as the data.
The correct description of the data by the simulation is checked
by independent measurements. Residual
differences between data and simulation, e.g. in the efficiencies 
of the lepton identification or of the trigger, are included in the
systematic uncertainties (Table~\ref{syserr}). 

The Monte Carlo generator used for inelastic $J/\psi$ production
 is EPJPSI \cite{epjpsi} which generates events according to the Colour
Singlet Model in leading order. In contrast to the standard version used 
previously~\cite{Adloff:1999zs}, the full dependence of the matrix 
element on $Q^2$ has been implemented~\cite{kruecker}. 
In order to achieve a good description of the data, the events are reweighted
in \qsq\ using a parametrisation of the measured \qsq\ distribution.
A systematic uncertainty of $\pm5\%$ is estimated for this procedure by 
repeating the analysis without this reweighting.

Diffractive production of \jpsi\ and \psits\ mesons is simulated using  
DIFFVM~\cite{diffvm} with parameters which have been tuned to HERA measurements.
Contributions from the production of $\bbbar$\ quark pairs with subsequent formation and 
decay of $b$-flavoured hadrons, \bdec, are simulated by the AROMA Monte Carlo 
program \cite{aroma}. The total AROMA cross section is normalised to the 
measured value of 16.2 nb~\cite{h1b}.
\subsection{Radiative Corrections}\label{radcor}
The measured cross sections are given in the QED Born approximation.
The effects of higher order processes, mainly initial state
radiation, are estimated using the HECTOR program \cite{hector}. 
With the requirement that $\sum (E-p_z)>40\mbox{~GeV}$ 
(see below) the radiative corrections amount to $-(4\dots5)\%$ and depend
only weakly  on $Q^2$ and $W$. 
A correction of $-(5\pm4)\%$ is applied.

%%%%%%%%%%%%%%%%%%%%%%%%%%%%%%%%%%%%%%%%%%%%%%%%%%%%%%%%%%%%%%%%%%%%%%%
%----------------------------------------------------------------------
%%%%%%%%%%%%%%%%%%%%%%%%%%%%%%%%%%%%%%%%%%%%%%%%%%%%%%%%%%%%%%%%%%%%%%%
\section{Data Analysis}
\subsection{Event Selection}
Events with $Q^2 > 2 \,\gevt$  are selected by requiring a scattered
lepton with a minimum energy deposition 
of $12\mbox{~GeV}$ in the electromagnetic calorimeter and a lepton 
scattering angle larger than $3^{\circ}$.
%Cuts are also applied to the cluster position and cluster shape in order to ensure high
%trigger efficiency and a good quality lepton measurement. 
The $z$ coordinate of the vertex position is determined for each event 
and required to lie in the beam interaction region.
In order to minimise the effects of QED radiation in the initial
state, the difference
between the total energy and the total longitudinal momentum $\sum (E-p_z)$ 
reconstructed in
the event is required to be larger than $40\mbox{~GeV}$. If no
particle, in particular no radiated photon, has escaped detection in the backward 
direction, the value of $\sum (E-p_z)$ is expected to be twice the incident lepton
energy,~i.e.~$55\mbox{~GeV}$. 

\begin{figure}[t!] \centering
\begin{picture}(18,8)  
\put(2.,5.6){a)}
\put(9.7,5.6){b)}
  \put(-1.,-0.9){\epsfig{file=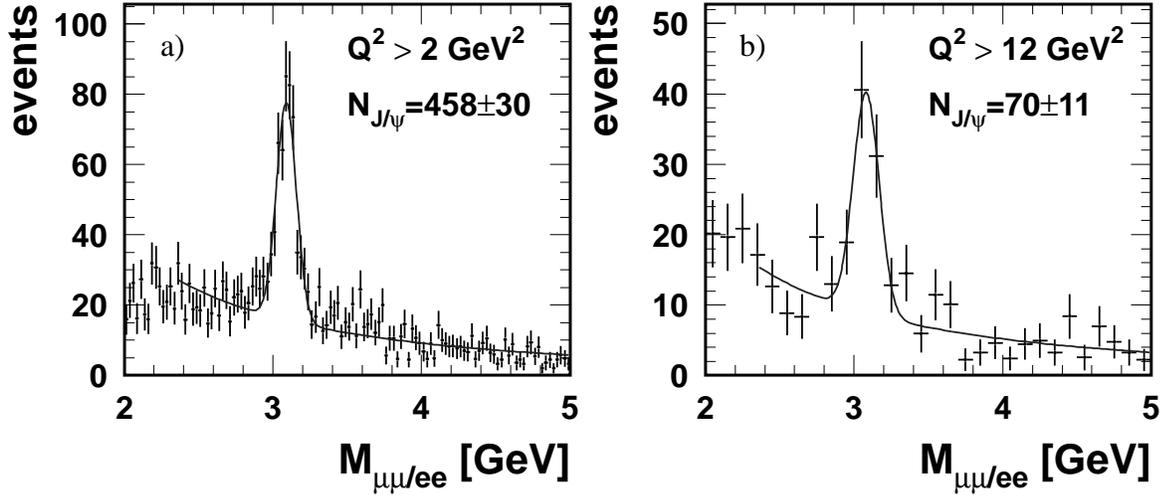,width=18cm}}
\end{picture}
\caption{Sum of di-muon and di-electron mass spectra for a) $Q^2>2\,\gevt$ 
and b) $Q^2>12\,\gevt$ after all selection cuts. The curves are the results 
of fits of Gaussian distributions for the signal   
%(convoluted with an exponential tail to account for energy loss in
%the case of di-electron decays) 
and a power law for the non-resonant background.}
\label{signalsi}
\end{figure}

The $J/\psi$ decay leptons are reconstructed as two 
oppositely charged particles  with transverse momenta $p_t >
0.8\mbox{~GeV}$ in the CTD. Both tracks have to be identified as  
muons with polar angles in the range $20^\circ < \theta < 160^\circ$
or as electrons in the range $30^\circ < \theta < 150^\circ$.
There is a considerable non-resonant background,  
mainly due to misidentified leptons (compare Fig.~\ref{signalsi}), 
in particular at low values of $z$. 
Therefore the number of \jpsi\ 
candidate events in a given analysis interval is extracted by fitting the mass 
distribution with   
a superposition of a Gaussian of fixed
width and position (determined by a fit to all data) 
to describe the signal and a power law component to describe the background.  
The number of signal events is then obtained by 
counting the number of lepton pairs in the interval
$2.85<M_{\mu\mu}<3.35\,\gev$ and subtracting the fitted amount of
background in this interval. This method was found to give stable and reliable results 
in most regions of phase space. The statistical error 
on the number of signal events is estimated from the statistical error on the
number of events (signal+background) in the mass interval.  
This method leads to a loss of events for the decay of 
the \jpsi\ to electrons due to radiation of the decay electrons in 
the material of the detector and due to decays $\mbox{$\jpsiw\ra e^+e^-\gamma$}$. 
A correction of $\sim 10\%$ is applied.
A systematic uncertainty of $3-7\%$ depending on kinematic variables is
estimated for the determination of the signal event numbers by changing the 
functional form for the background.

After the cuts described above the main background is due to the diffractive production of 
 \jpsi\ mesons which is concentrated at high $z$ values. 
Diffractive \jpsi\ contributions can experimentally be suppressed in several
ways. Previously, inelastic events were selected by requiring the  
hadronic system $X$, which is  produced together with the \jpsi\ meson, to have a 
high mass~\cite{Adloff:1999zs}. In the present analysis, 
a selection cut $z<0.9$ is used to suppress
 diffractive elastic and proton dissociative events. This range 
corresponds to the region of validity of the theoretical calculations 
which are used for comparison.
A further cut is applied, \ptst$>1\,\gevt$, where \ptstar\ is the 
transverse momentum of the \jpsi\ in the photon proton \cms\ frame.
After this requirement, the background from diffractive 
\jpsi\ meson production is estimated to be less than 2\% and is neglected.

\subsection{Contributions from \boldmath$b$ and $\psi(2S)$ Decays}
After the cuts described in the previous section, the 
\jpsi\ sample is dominated by `direct' inelastic \jpsi\ production, 
in which the \jpsi\ is directly produced from the \ccbar\ pair 
in the process $\gamma^*g\ra\ccbar\,(g)$.
However there remain contributions from both the diffractive and inelastic 
production of \psits\ mesons 
and the production of $b$ flavoured hadrons with subsequent decays to
states involving  
\jpsi\ mesons.
Diffractive \psits\ events are expected mainly at high $z$ values 
while contributions from  \bdec\ are expected at low $z$ values.
With the cut on \ptst$>1\,\gevt$ the remaining background from \psits\ is 
mainly
due to diffractive events in which the proton dissociates and is estimated to be 
between 6\% and 20\% in the highest $z$ bin, $\mbox{$0.75<z<0.9$}$, corresponding to 
$2-10\%$ in the total sample. The lower estimate (6\%) is based on  
a Monte Carlo simulation of diffractive \psits\ production 
(the simulated contribution is shown in Figs.~\ref{dataMCinela}c) and
e).  Since diffractive \psits\ production has not been measured in the 
present kinematic range we consider this to be a crude estimate. An analysis of
the present data, where  events with less than five particles are selected
as candidates for $\psits\ra\jpsiw\,\pi^+\pi^-$, 
yields an estimate of 20\% in the highest $z$ bin. 
No correction is applied to the data, since the dependence on the kinematic 
variables, in particular on the transverse momentum of the \jpsi\ meson, is  
poorly known. %For this reason no systematic error is taken account in the 
%differential distributions.

The contribution from \bdec, which is expected at low values of $z$,
is estimated from a Monte Carlo simulation of $\bbbar$ 
production~\cite{aroma} using the measured $b$ cross section 
 \cite{h1b}. It is estimated to be 17\% in the lowest $z$ bin 
($0.3<z<0.45$, compare Fig.\ref{dataMCinela}c) corresponding to 5\% 
in the total sample. Again, no correction is applied to the data due to 
the poorly known dependences on the kinematic parameters.  

Inelastic production of \psits\ mesons with subsequent decays 
\psits$\ra \jpsiw+X$ give a further contribution 
which at present cannot be distinguished experimentally. It is expected to 
contribute over the whole $z$ range and its dependence on the kinematic
variables is likely to be similar to that of direct \jpsi\ production.
It can thus be considered as a normalisation uncertainty. In the
photoproduction limit this contribution is estimated to be 
$\sim 15\%$\cite{nrqcd}.

Summarising, the measured cross sections  contain in addition to
direct inelastic \jpsi\ mesons, contributions from diffractive \psits\ events 
and $b$ decays 
which may amount to as much as 17\% in total. The distributions
in the variables studied have not been measured, but are expected to be quite different 
for the two processes and different from those of the direct inelastic \jpsi\ events. 
No correction or systematic error is applied.
Inelastic \psits\ events on the other hand are expected to
have similar distributions to the inelastic \jpsi\ events themselves; their
contribution may be of the order of 15\% and can be regarded as a normalisation 
uncertainty.

\section{Results}
Differential cross sections are determined in the kinematic region
$ 2 <Q^2 <100\,\mbox{~GeV}^2$ ($\langle Q^2\rangle= 10.6\mbox{~GeV}^2$), 
$50<W<225\,\mbox{~GeV}$, $\ptstw>1\,\gevt$ and  $0.3<z<0.9$. 
A second set of differential cross sections is determined for a subset
with $Q^2 > 12 \mbox{~GeV}^2$ and, in order to match the \qsq\ range,
with $p_{t,\psi}^2 > 6.4 \mbox{~GeV}^2$. The  average
 value of \qsq\ in this sample is $\langle Q^2\rangle=30.9\,\gevt$.
The distribution of the invariant mass of the two leptons after 
all selection cuts is shown in Fig.~\ref{signalsi}. 
The total number of signal events is $458\pm30$ of which $70\pm11$ 
 are at \qsq$>12\,\gevt$ and $p_{t,\psi}^2 > 6.4 \mbox{~GeV}^2$.

\begin{figure}[p] 
\centering
\begin{picture}(16.0,15.5)
  \put(-0.1,-0.9){\epsfig{file=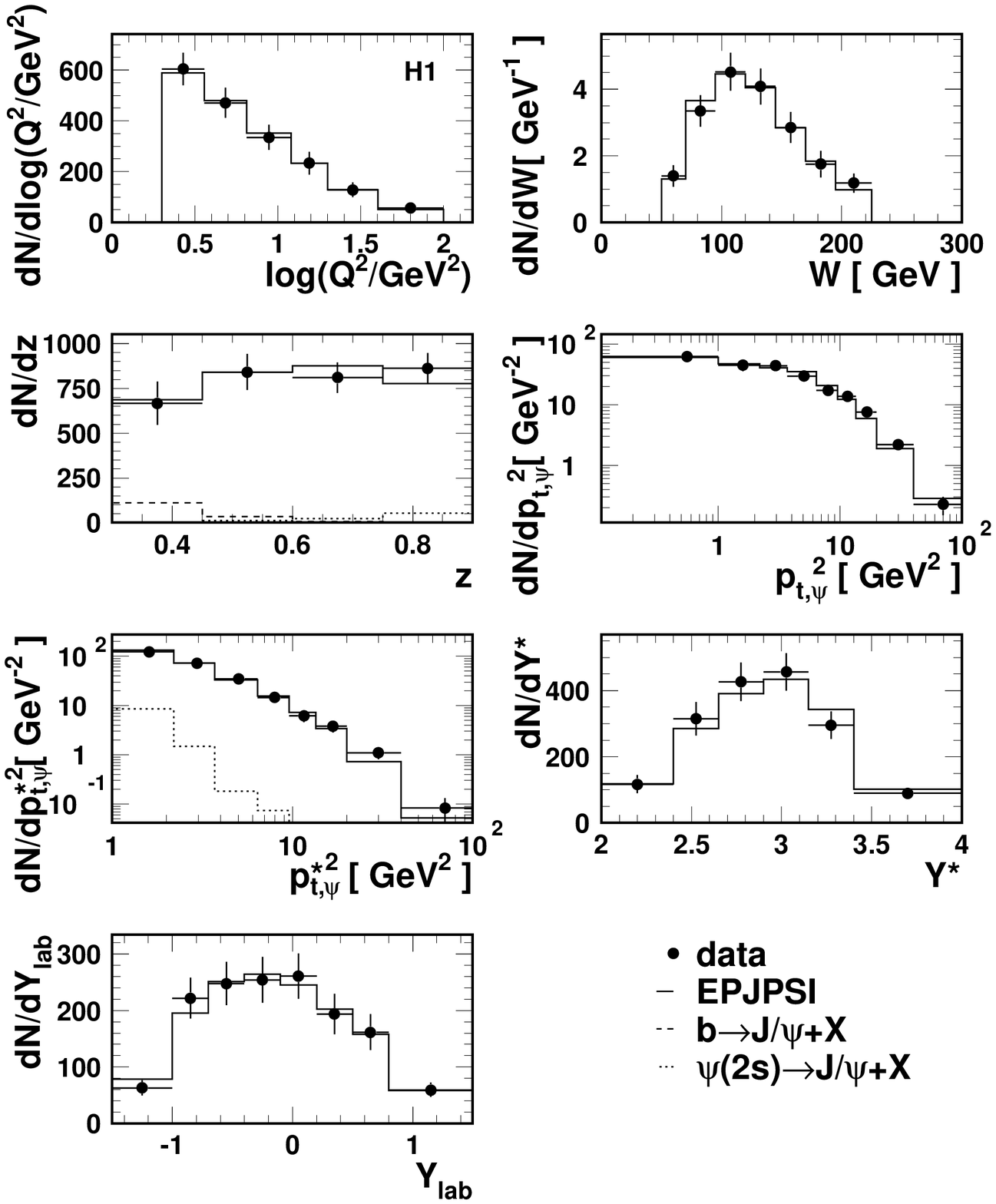,width=17.5cm}} 
\put(2.5,16.6){a)}
\put(2.5,12.2){c)}
 \put(6.5,7.8){e)} 
\put(2.5,3.4){g)}
\put(14.,16.6){b)}
\put(14.,12.2){d)}
\put(14.,7.8){f)}
%\put(12.5,2.4){\bf\large +X}
\end{picture}
\caption{Comparison between data and Monte Carlo simulations for 
         $ep \rightarrow e\:J/\psi\: X$ in the region $50<W<225\,\gev$, 
         $\ptstw>1\,\gevt$ and $0.3<z<0.9$
         after all selection cuts and after subtraction of non-resonant background.  
  Distributions of a) $Q^2$, b) $W$, c) $z$, d)
  $p_{t,\psi}^2$, e) $p_{t,\psi}^{*2}$, 
  f)  $Y^{*}$ and g) $Y_{lab}$ are shown.   
The error bars on the data points are statistical only. The full histograms 
show the inelastic \jpsi\ events from EPJPSI after weighting the events according to 
a parametrisation of the \qsq\ 
dependence of the data and normalising over the $z$ range measured. The estimated 
contribution from $b$ decays (AROMA, dashed) is also shown in c) and from 
diffractive \psits(EPJPSI, dotted) in c) and e). These
contributions are not included in the full histograms.}
\label{dataMCinela}
\end{figure}

Comparisons between the data and the Monte Carlo simulation (EPJPSI), which is used 
to correct for detector effects, are shown in Fig.~\ref{dataMCinela}. The simulations 
take into account the two lepton proton centre of mass energies according to the
luminosity.  The simulation is normalised to the data in the 
interval $0.3<z<0.9$ after reweighting the events in $Q^2$ 
and then describes all other distributions well. Monte Carlo estimates
of contributions from \bdec\
and diffractive \psits\ production are indicated in the $z$ and \ptst\ 
distributions (Fig.~\ref{dataMCinela}c and e). 
%Although these contributions may be significant in certain regions of phase space
%amount up to 17\%
%they are neglected in the following (compare also previous section).
%%%%%%%%%%%%%%%%%%%%%%%%%%%%%%%%%%%%%%%%%%%%%%%%%%%%%%%%%%%%

The systematic uncertainties in this analysis are typically $15-17\%$ and amount 
to 21\% in single bins at low $z$ and $W$. For the double differential cross 
sections the corresponding error estimate is 21\%. The systematic errors 
are dominated by uncertainties in estimating the number of events in
regions of high non-resonant background, by the uncertainty in the Monte
Carlo calculation used for acceptance and efficiency corrections, and by 
uncertainties in
efficiencies for lepton identification and triggering. A list
is given in Table \ref{syserr}. 

\begin{table}[h!] \centering
\begin{tabular}{lr} \hline
Source & Uncertainty [\%]\\[.1em]
\hline
Reconstruction of scattered lepton: angle  & 5\\ 
\hspace{5.9cm} ~energy & 5\\ 
Reconstruction of decay leptons: track, vertex&4\\
\hspace{5.5cm} ~identification & 6\\
Number of signal events & 3--15\\
Trigger & 7.3\\
Monte Carlo model & 5 \\
Hadronic energy scale& 4\\
Radiative corrections & 4\\
Integrated luminosity & 1.5\\
Decay branching ratio & 2\\
\hline
Sum & 15--21\\
\hline%\hline
%Background & $<$9-17\\
%\hline
\end{tabular}
\caption{Summary of systematic errors for the single differential 
\jpsi\ production cross sections. 
The error on the number of events depends on
$z$ and \ptst. %In single intervals, e.g. at low $z$ or $W$ it is
%higher than the quoted range ($\sim15\%$). 
The total
error is the sum of the contributions added in quadrature.}
\label{syserr}
\end{table}

\subsection{Differential Cross Sections}

\begin{figure}[h] \centering
\begin{picture}(17.,9.5) 
  \put(-3.0,-0.9){\epsfig{file=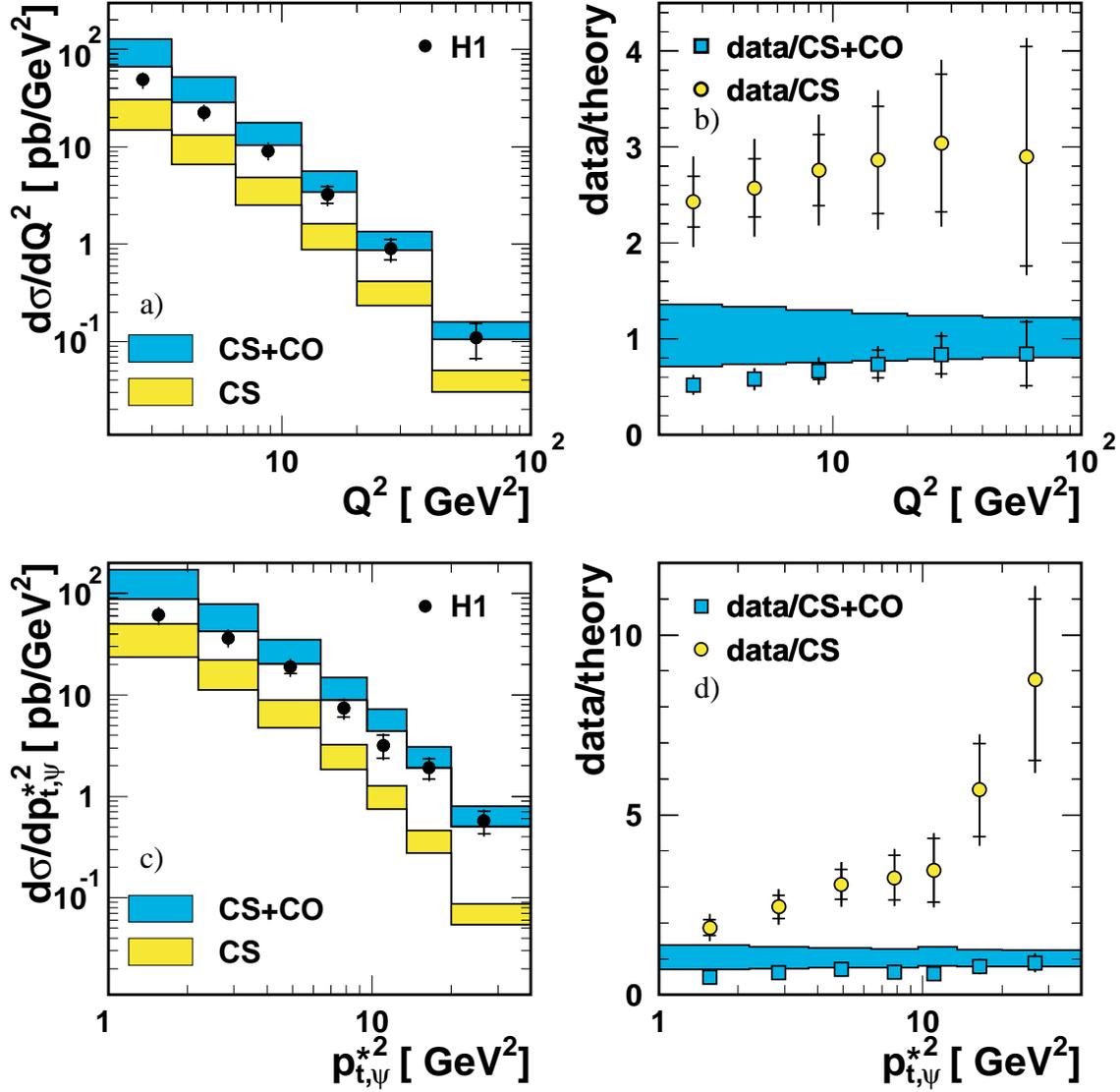,width=19.5cm}}
\put(2.,10.5){a)}
\put(2.,3.){c)}
\put(9.5,13.){b)}
\put(9.5,5.3){d)}
\end{picture}
\caption{Differential cross sections  a) $d\sigma / dQ^2$ and c) \dsdpts\ 
         for the inelastic process 
         $ep \rightarrow e\:J/\psi\: X$ in the region $50<W<225\,\gev$, 
         \qsq$>2\,\gevt$, $\ptstw>1\,\gevt$ and $0.3<z<0.9$. 
         The NRQCD calculation is shown for comparison (CO+CS, light 
         band) and the \colsing\ contribution
         separately (CS, dark band).
         In b) and d) the ratio of data/theory is shown for the two cases. The
         theoretical uncertainty in the full calculation (CS+CO) is shown as a 
         band around $1$. 
         The inner error 
         bars of the data are statistical, the outer  error bars contain 
         statistical and 
         systematic uncertainties added in  quadrature.}
\label{lqq2}
\end{figure}

\begin{figure}[p] \centering
\begin{picture}(15.8,18.)
  \put(-1.0,-0.9){\epsfig{file=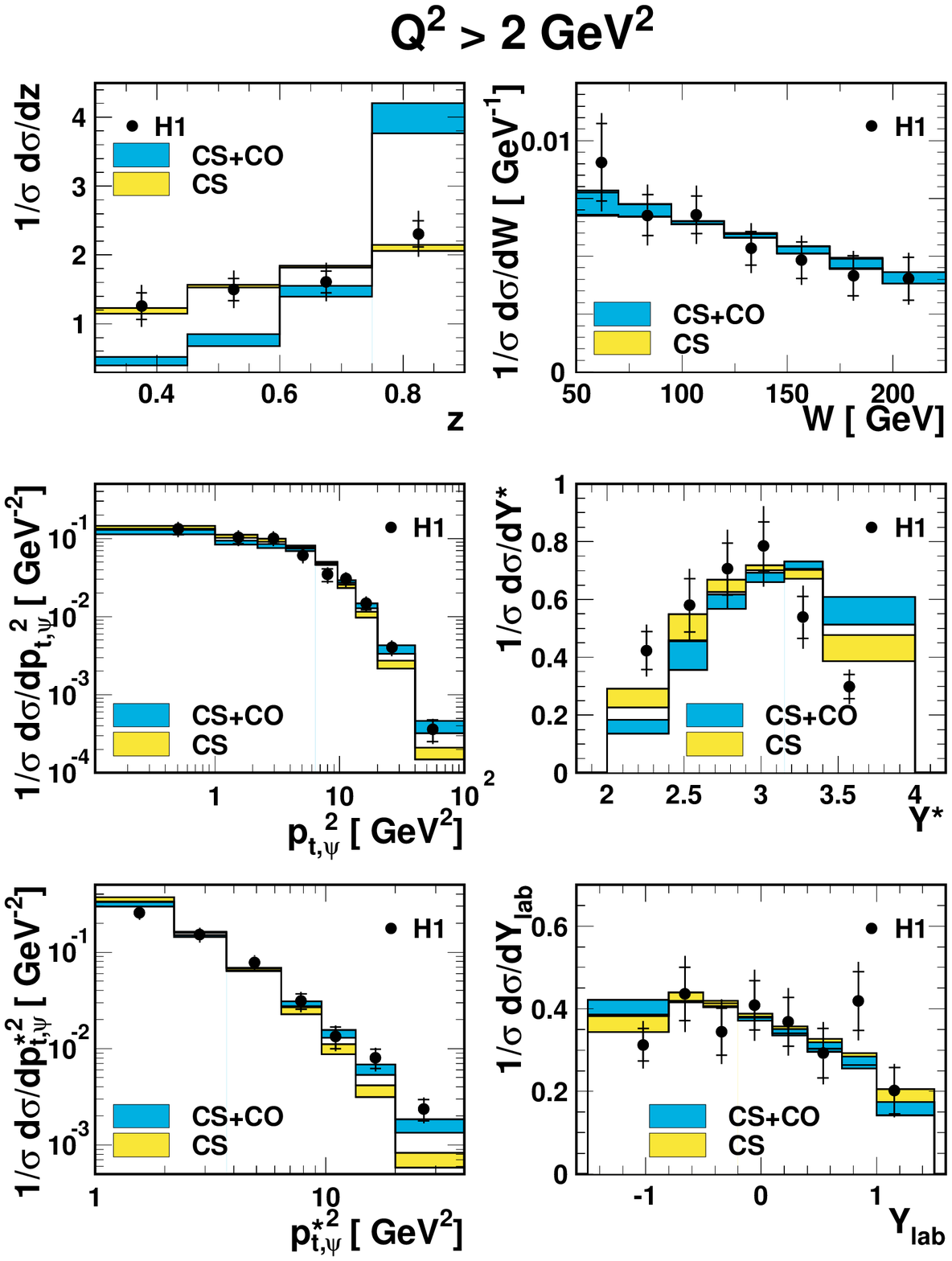,width=17.5cm}}
%\put(6.5,19){\large\boldmath\bf$Q^2>2\,\gevt$}
\put(2.,16.){a)}
\put(2.,10.2){c)}
 \put(2.,4){e)} 
\put(9.5,17.7){b)}
\put(9.5,11.4){d)}
\put(9.5,5){f)}
\end{picture}
\caption{Normalised differential cross sections for the inelastic process 
$ep \rightarrow e\:J/\psi\: X$ in the kinematic region  
$2 < Q^2 < 100\mbox{~GeV}^2$, $50 < W < 225\mbox{~GeV}$, $\ptstw>1\,\gevt$ and $0.3<z<0.9$. 
a) $1/\sigma\,d\sigma / dz$, b) $1/\sigma\,d\sigma / dW$,
c) $1/\sigma\,d\sigma / dp_{t,\psi}^2$, d)$1/\sigma\,d\sigma /dY^\ast$ 
e) $1/\sigma\,d\sigma / dp_{t,\psi}^{*2}$ and f) $1/\sigma\,d\sigma /dY_{lab}$.    
The inner 
error bars are statistical, the outer  error bars contain statistical and 
systematic uncertainties added in  quadrature. 
The histograms show calculations for inelastic $J/\psi$ production within 
the NRQCD factorisation approach\protect\cite{kniehl} which have been normalised 
to the integrated cross section. The light 
band represents the sum of CS and CO contributions and the dark 
band the CS contribution alone (both are separately
normalised). The error bands reflect the theoretical uncertainties (see text).}
\label{lqrest}
\end{figure}

The differential cross sections for inelastic $J/\psi$ production
are displayed in Fig.~\ref{lqq2}  as functions of $Q^2$ and \ptst. 
In Fig.~\ref{lqrest} normalised differential cross sections are shown
as functions of the 
variables  $W$, $z$, $p_{t,\psi}^2$, $p_{t,\psi}^{*2}$, $Y_{lab}$ and $Y^*$. 
The data points are plotted at the mean value of the data in each interval. 
The differential 
cross sections are also listed in Tables~\ref{lq_tab1} and \ref{lq_tab2}.
The results of the calculations 
by Kniehl and Zwirner\cite{kniehl}, who applied the NRQCD %/factorization
approach to electroproduction of $J/\psi$ mesons, are
shown for comparison. These calculations only include $2\ra
2$ contributions, which is appropriate for $z<0.9$.
For easier comparison of shapes the data and
the calculation in Fig.~\ref{lqrest} have been normalised to the  
integrated cross sections in the measured range for each distribution.
 
\begin{figure}[p] \centering
\begin{picture}(15.8,15.)
\put(-0.0,-0.9){\epsfig{file=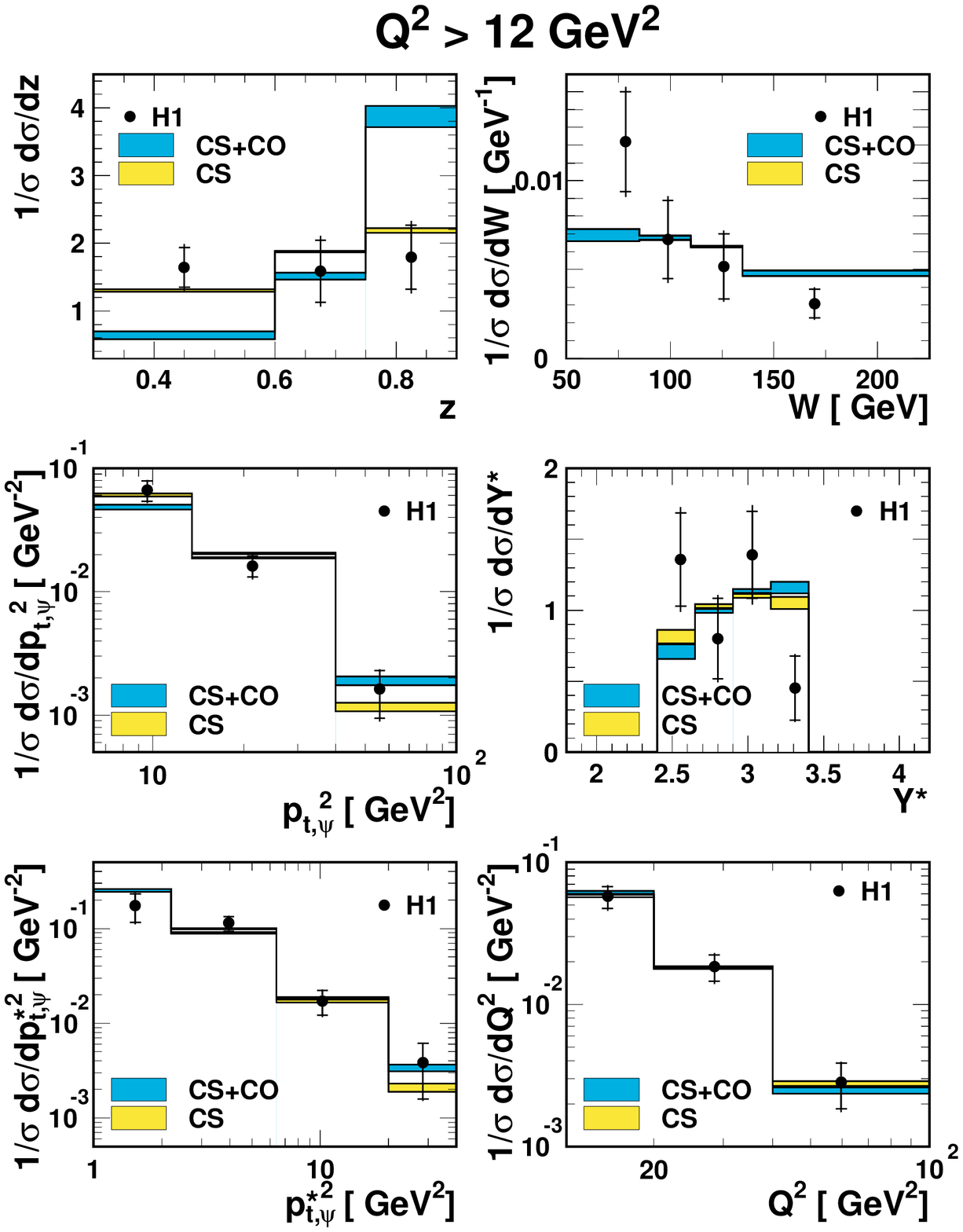,width=15.5cm}}
%\put(6.5,17){\large\boldmath\bf$Q^2>12\,\gevt$}
\put(4.3,15.6){a)}
\put(4.3,10.){c)}
 \put(4.3,4.4){e)} 
\put(10.7,15.6){b)}
\put(10.7,10.){d)}
\put(10.7,4.4){f)}
\end{picture}
\caption{Normalised differential cross sections for the inelastic process 
$ep \rightarrow e\:J/\psi\: X$ in the kinematic region  
$12 < Q^2 < 100\mbox{~GeV}^2$, $50 < W < 225\mbox{~GeV}$, $\ptt>6.4\,\gevt$, 
$\ptstw>1\,\gevt$ and $0.3<z<0.9$. 
a) $1/\sigma\,d\sigma / dz$, b) $1/\sigma\,d\sigma / dW$,
c) $1/\sigma\,d\sigma / dp_{t,\psi}^2$, d) $1/\sigma\,d\sigma /dY^\ast$ 
e) $1/\sigma\,d\sigma / dp_{t,\psi}^{*2}$ and f) $1/\sigma\,d\sigma / dQ^2$.    
The inner error bars of the data points are statistical, the outer  error bars contain 
statistical and systematic uncertainties added in  quadrature. 
The histograms show calculations for inelastic $J/\psi$ production within 
the NRQCD factorisation approach \protect\cite{kniehl}. The light 
band represents the sum of CS and CO contributions and the dark 
band the CS contribution alone (both are separately
normalised). The error bands reflect
the theoretical uncertainties (see text).}
\label{hq}
\end{figure}

The NRQCD calculations shown in the figures include
the contributions from the colour octet states\footnote{Spectroscopic notation 
is used: $^{2S+1}L_J$ where $S$, $L$ and $J$ denote 
 the spin, orbital and total angular momenta of the $c\bar{c}$ system.}
 $^{3}\!S_1$, $^{3}\!P_{J=0,1,2}$, $^{1}\!S_0$ 
as well as from the colour singlet state  $^{3}\!S_1$ (labelled ``CO+CS''). 
The contribution of the \colsing\ state is also shown separately
(labelled ``CS''). 
The calculations depend on a number of parameters. The values used for the 
non-perturbative long  range transition matrix elements (LDMEs) were determined from 
the distribution of transverse momenta of \jpsi\ mesons produced in \ppbar\ 
collisions~\cite{Braaten:2000qk}\footnote{\label{fn}
The extracted values for the LDMEs depend on the parton density 
distributions. For the set MRST98LO \cite{PDF} the values are, in the
notation of \cite{kniehl},
$\langle {\cal O}^{J/\psi}[^3\!S_1^{(1)}]\rangle=1.3\pm 0.1 \,$GeV$^3$, 
$\langle {\cal O}^{J/\psi}[^3\!S_1^{(8)}]\rangle=(4.4\pm0.7)\cdot10^{-3}\,$GeV$^3$ and
$M_{3.4}^{J/\psi}=(8.7\pm 0.9)\cdot10^{-2}\,$GeV$^3$.
$M_{3.4}^{J/\psi}$ is a linear combination of two elements,
$M_r^{J/\psi}=\langle{\cal O}^{J/\psi}[^1\!S_0^{(8)}]\rangle+r\langle {\cal O}^{J/\psi}[^3\!P_0^{(8)}] 
\rangle/m_c^2$, with $r=3.4$.}.
The bands in Figs.~\ref{lqq2}, \ref{lqrest} and \ref{hq} 
indicate the uncertainty in the theoretical calculation \cite{zwirner}. 
They cover a charm quark mass of $m_c=1.5\pm0.1\,\gev$,
variation of renormalisation and factorisation scales by factors 0.5
and 2, the errors of the LDMEs as well as  the
case that either of the two parts of $M_{r}^{J/\psi}$ (see footnote
\ref{fn}) does not contribute.  Furthermore, the effect 
of using the CTEQ5M \cite{cteq} set of parton density functions
instead of MRST98LO \cite{PDF} is included.

%%%%%%%%%%%%%%%%%%%%%%%%%%%%%%%%%%%%%%%%%%%%%%%%%%%%%%%%%%%%
The colour octet contribution dominates the predicted cross section
for all values of $Q^2$ and \ptst\  
(Fig.~\ref{lqq2}a and c). In order to facilitate the comparison with the data, 
the ratio  data/theory is shown on a linear scale in Fig.~\ref{lqq2}b and d 
together with a band indicating the 
uncertainty in the NRQCD calculation with CO+CS contributions.
The NRQCD calculation overshoots the data by about a factor of 2 at low
\qsq, which is at the limit of the theoretical and experimental error.
%absolute values of data/theory are approximately  $\sim0.52$ at
%low \qsq\ increasing to $\sim 0.85$ for high \qsq. 
The agreement between the data  and the theory improves towards higher \qsq\ where the
theoretical uncertainties diminish.
For \ptst\ similar agreement between data and NRQCD calculation is observed. 
Compared with the \colsing\ contribution alone, the data exceed the
calculations by a factor $\sim 2.7$, approximately 
independent of \qsq, while for \ptst\ the ratio increases towards higher values of 
\ptst.

In Fig.~\ref{lqrest} the measured and the theoretical differential cross 
sections are normalised to the integrated cross sections in the measured range 
for each distribution.
The $W$ and the $Y_{lab}$ distributions 
(Fig.~\ref{lqrest}b and f) are reasonably well described in shape by the full 
NRQCD calculation 
and also by the \colsing\ contribution alone, whereas neither fully describes
the $Y^*$ distribution. The $z$ distribution  is very poorly 
 described by the full calculation including \coloct\ contributions, while 
the \colsing\ contribution  alone reproduces the shape of the data rather well.   
%This has also been found in the photoproduction limit \cite{H1962,Zeus972,katja}. 
A similar discrepancy between data and NRQCD calculations was observed
at high $z$ values in the photoproduction limit \cite{H1962,Zeus972,katja}. 
It may be due to phase space limitations at high $z$ for the emission 
of soft gluons in
the transition from the \coloct\ \ccbar\ pair to the \jpsi\ meson, which are not taken into account in the calculation. In photoproduction,  
the rapid rise of the \coloct\ contributions towards high $z$ values was
shown to be damped after resummation of the NRQCD expansion \cite{wolf,katja}. 
The shapes of the \ptt\ and \ptst\ distributions
(Fig.~\ref{lqrest}c and e) are rather well described by including CO+CS contributions 
while the CS contribution alone decreases too rapidly towards high values of
\ptt\ or \ptst. Note, however, that higher orders are
expected to contribute  significantly at high values of $p_{t,\psi}$  as
observed in next-to-leading order CSM calculations in the photoproduction 
limit\cite{Kraemer}.

%%%%%%%%%%%%%%%%%%%%%%%%%%%%%%%%%%%%%%%%%%%%%%%%%%%%%%%%%%%%
At higher \qsq\ values the theoretical uncertainties of the NRQCD calculation 
decrease (see Fig.~\ref{lqq2}b). It is therefore interesting to compare   
data and theory at higher \qsq. The results for \qsq$>12\,\gevt$  
(with \ptt$>6.4\,\gevt$) are given in Fig.~\ref{hq} and Table~\ref{hq_tab}. 
The requirement \ptst$>1\,\gevt$ is retained. 
The average $\langle Q^2\rangle = 30.9\mbox{~GeV}^2$  is 
larger than the squared mass of the \jpsi\ meson, $m_{\jpsiw}^2$. 
 The statistical precision of these data is limited and no 
substantial change in the comparison of data and theory is seen
compared to Fig.~\ref{lqrest}.

%%%%%%%%%%%%%%%%%%%%%%%%%%%%%%%%%%%%%%%%%%%%%%%%%%%%%%%%%%%%
\subsection{Double Differential Cross Sections}
\begin{figure}[h] \centering
\begin{picture}(10.8,8.5)
\put(-3.5,-1.){\epsfig{file=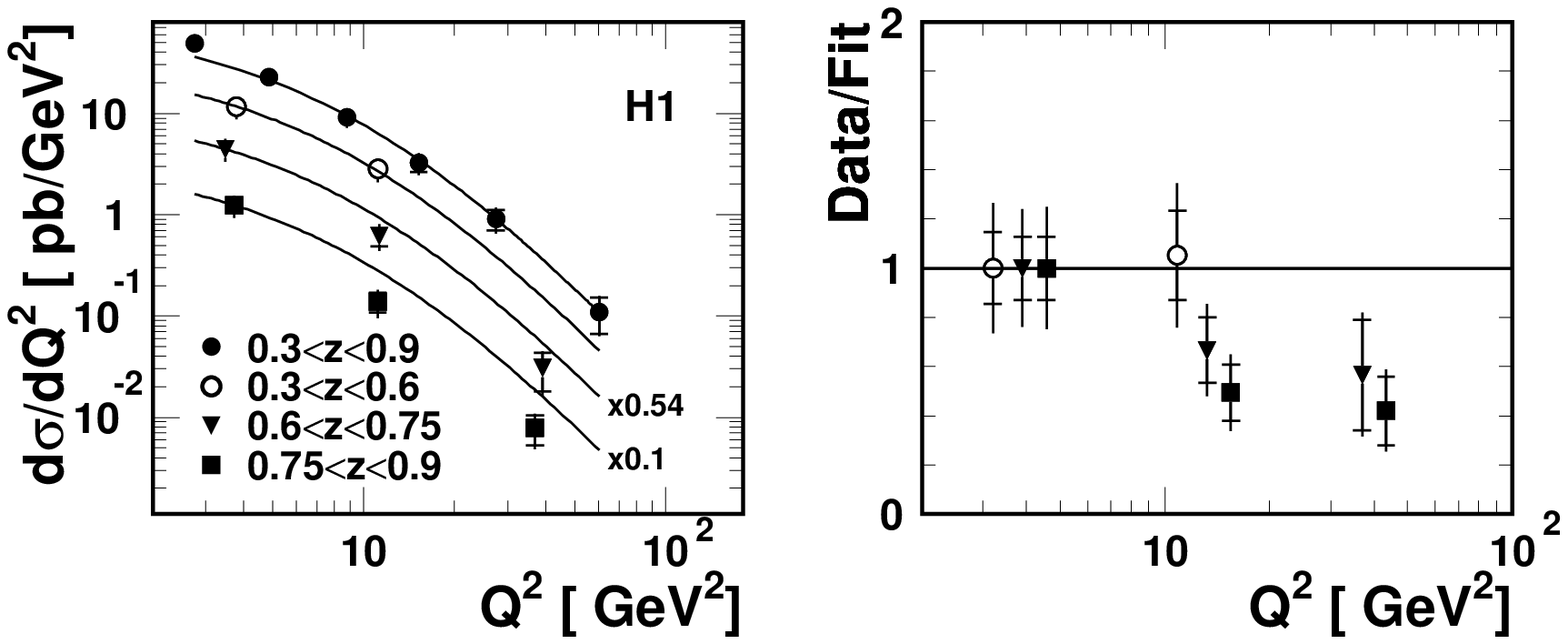,width=18cm}}
\put(-3.5,6.){\epsfig{file=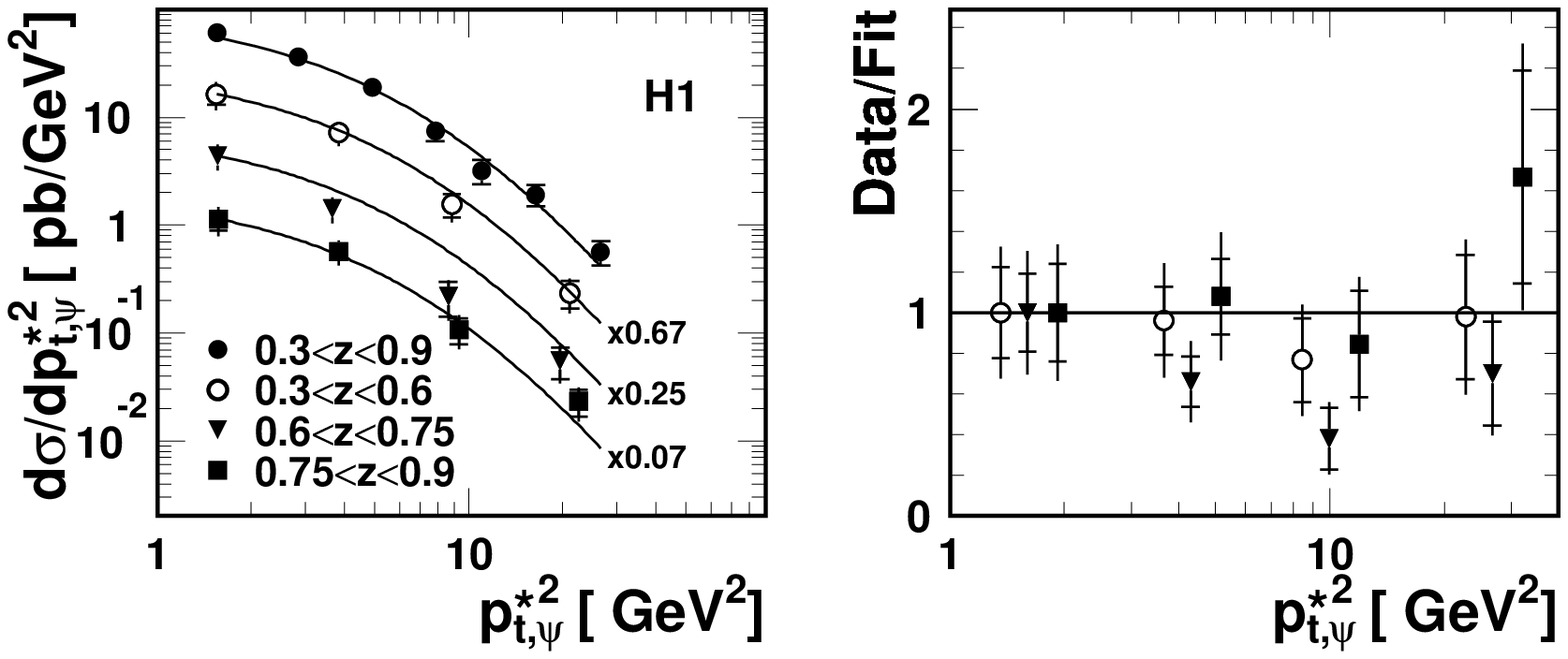,width=18cm}} 
%\put(-3.5,-1.){\epsfig{file=fig2d_zq.final.sysneu.eps,width=18cm}}
%\put(-3.5,6.){\epsfig{file=fig2d_zps.final.sysneu.eps,width=18cm}} 
\put(1.7,12.6){a)}
\put(1.7,5.6){c)}
\put(7.2,12.6){b)}
\put(7.2,5.6){d)}
\end{picture}
\caption{Differential cross sections for $e\,p\rightarrow e\,J/\psi \,X$
in three $z$ intervals and in the full $z$ range.
a) $d\sigma /dp_{t,\psi}^{*2}$ and c) $d\sigma / dQ^2$
for low ($0.3<z<0.6$, open points), medium ($0.6<z<0.75$, triangles) and
high ($0.75<z<0.9$, squares) values of $z$ in comparison with the
results for the full $z$ region (full points). 
The inner error bars indicate the
 statistical uncertainty, while the outer error bars show the 
statistical and systematic uncertainties added in quadrature.
For clarity, the data have been scaled by the factors indicated. 
The data in the complete $z$ range are parametrised by fits of the form 
(\qsq$+m_{\jpsiw}^2)^{-n}$ and (\ptst$+m_{\jpsiw}^2)^{-m}$. The same 
parametrisations are
also shown for the data in the three $z$ bins after normalising them 
to the data at low \qsq\ or \ptst.
In b) and d) the ratio of the data to these parametrisations is shown on a linear
scale using the same symbols as in a) and c). Note that  for clarity the points have 
been shifted in \qsq\ and \ptst.}
\label{ddif}
\end{figure}

In the calculations the relative contributions of the \coloct\ states to 
the cross sections vary with $z$, \qsq\ and \ptst\ (compare
Figs.~\ref{lqq2}, \ref{lqrest}a and c). Therefore, differential
cross sections $d\sigma/dp_{t,\psi}^{\ast 2}$ and $d\sigma/d$\qsq\ are determined% 
\footnote{The corresponding double differential cross sections 
$d\sigma/dp_{t,\psi}^{\ast 2} dz$
and  $d\sigma/d$\qsq $dz$\ are listed in Table~\ref{2d_tab}.} in three intervals 
of $z$ and 
compared with that for the whole $z$ range in Fig.~\ref{ddif}.
The dependence on \qsq\ and \ptst\ 
is seen to be similar in the three $z$ regions. 
In order to make a quantitive comparison, the differential cross sections for 
the whole $z$  range are fitted with functions $\propto(\qsq+m_{\jpsiw}^2)^{-n}$ or
$\propto(\ptstw+m_{\jpsiw}^2)^{-m}$ yielding ($n=3.36\pm0.53$) and  
($m=4.15\pm0.50$), respectively, where total experimental errors are given.  
The results of these same fits are then 
compared with the data in the three $z$ intervals after normalising the curves at 
low \qsq\ or \ptst.
In Figs. \ref{ddif}b 
and d the ratio of the data over the scaled fit is shown.
The data in the three $z$ bins are reasonably described by the
same functional form although there is an indication of a faster fall with 
\qsq\ at high $z$ than in the total $z$ range. 
In view of the contributions at high $z$ from diffractive
\psits\ production, which are expected to have a different dependence
on \qsq, firm conclusions cannot be drawn. 
The observed dependence on \ptst\ is within errors the same as that observed in 
photoproduction ($m\approx4.6\pm0.1$)\cite{katja}. 
%In conclusion, within present errors neither the
%dependence on \qsq\ nor on \ptst\ exhibit a significant dependence on $z$.
%

%%%%%%%%%%%%%%%%%%%%%%%%%%%%%%%%%%%%%%%%%%%%%%%%%%%%%%%%%%%%

\subsection{\boldmath \gstp\ Cross Sections and Comparison to Photoproduction}

For comparison with results in the photoproduction limit
the cross section for $\mbox{$\gamma^* p \rightarrow J/\psi\,X$}$ as a 
function of $W$ 
is calculated by dividing the $ep$ cross section by the photon flux integrated 
over the analysis intervals \cite{epa}. 

The total cross section $\sigma(\gamma^* p\ra \jpsiw\,X)$ is shown as a function of $W$
 in Fig.~\ref{inclw} and listed in Table~\ref{gp_tab1}
 for the present data ($\langle Q^2\rangle = 10.6\mbox{~GeV}^2$).
It is compared with the cross section in the photoproduction limit 
($\langle Q^2\rangle \sim0.05\,\mbox{~GeV}^2$) in an otherwise similar 
kinematic range\cite{katja}.
%The $W$ dependence is seen to be very similar for the two datasets.
Parametrising the cross section as $(W/W_0)^{\delta}$ yields  
%$\langle Q^2\rangle = 9.6\mbox{~GeV}^2$ a value 
$\delta = 0.65 \pm 0.25$ for the present data, where the total
experimental error was used in the fit. The value is consistent 
with that obtained in photoproduction ($0.49\pm0.16$~\cite{katja}). 
%Wert vor Aenderung der sys Fehler: ($0.51\pm0.14$~\cite{katja}). 
% $\sim 0.52\pm0.10$ (only statistical errors) 
%for \qsq$<1\,\gevt$ \cite{katja}. 
The $W$ dependences are expected to be similar since they reflect the $x$ 
dependence of the gluon distribution with a scale 
$\sim \qsq+m^2_{\jpsiw}$. 

\subsection{Decay Angular Distributions}

\begin{figure} \centering
\begin{picture}(10.8,8.5)
%\put(-3.5,-0.7){\epsfig{file=figlq_q2gp.final.eps,width=10.5cm}}
\put(0.0,-0.7){\epsfig{file=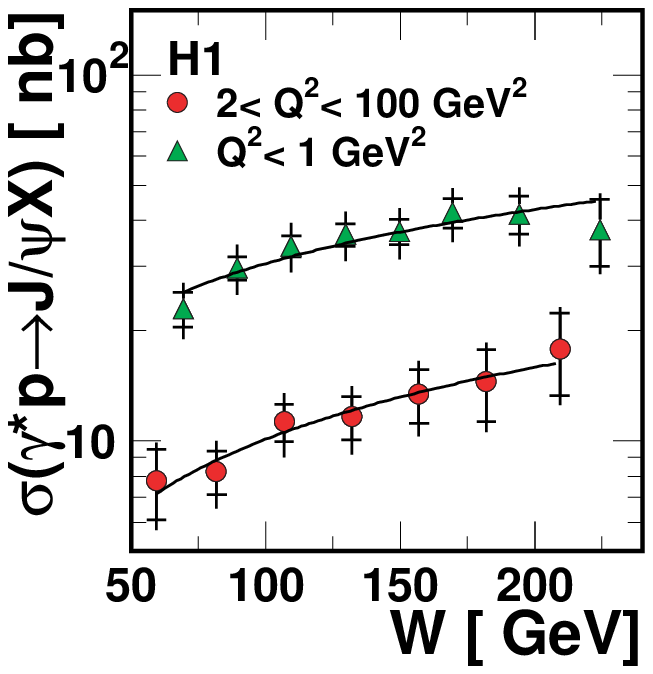,width=10.5cm}} 
%\put(-0.6,7.8){a)}
%\put(7.,7.8){b)}
\end{picture}
\caption{Total cross section for $\gamma^{\ast}\,p\rightarrow J/\psi \,X$ at  
$\langle Q^2\rangle = 10.6\mbox{~GeV}^2$ in the range 
$0.3 < z < 0.9$ as a function of $W$.
Photoproduction data \cite{katja} with similar cuts in $z$ and
\ptt\ are included for comparison. The curves  are the results of fits
with the function $(W/W_0)^\delta$.
The inner error bars on the points indicate the
 statistical uncertainty, while the outer error bars show the 
statistical and systematic uncertainties added in quadrature.}
\label{inclw}
%\end{figure}

%\begin{figure}[h] \centering
\begin{picture}(10.2,8)
  \put(-4.5,-1.0){\epsfig{file=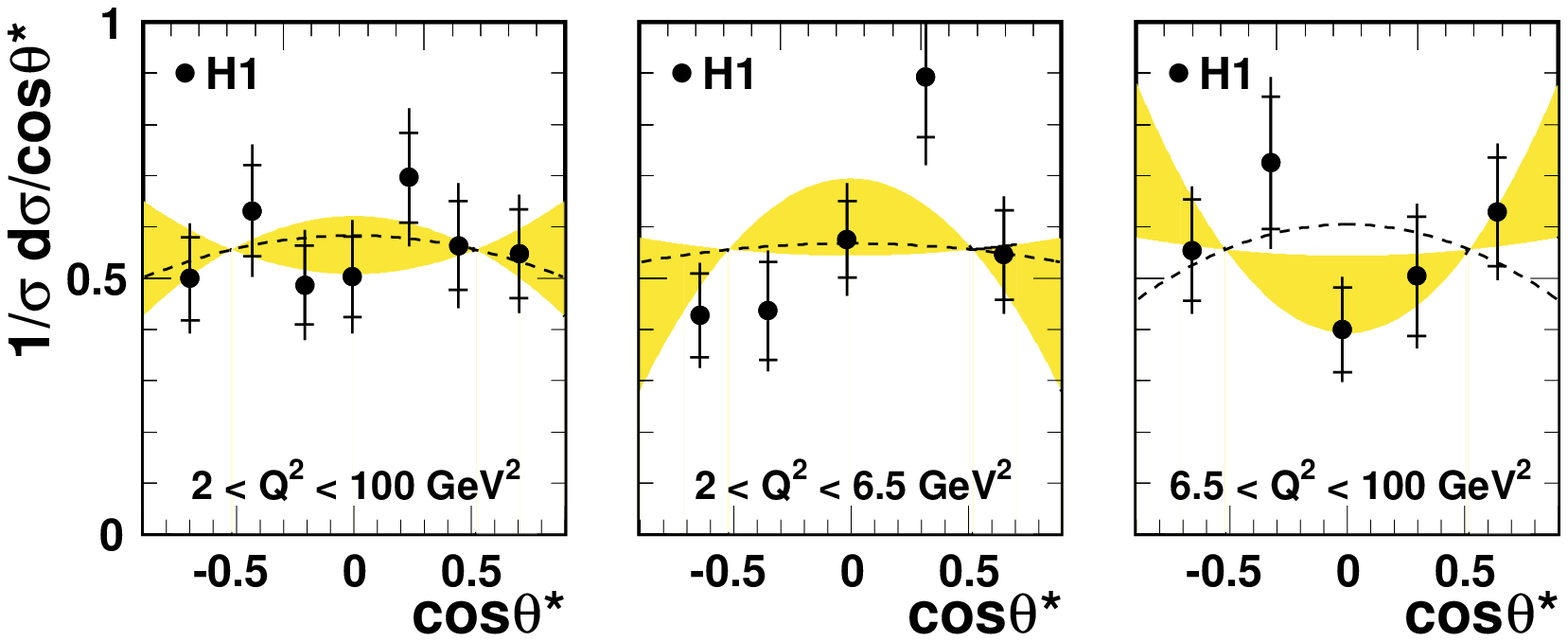,width=19.5cm}} 
% \put(-3.75,-1.0){\epsfig{file=figlqnor_costhfinal.neuzw.eps,width=18cm}} 
\put(0.4,6.2){a)}% \boldmath$2<\qsq<100\,\gevt$}
\put(5.55,6.2){b)}% \boldmath$2<\qsq<6.5\,\gevt$}
\put(10.6,6.2){c)}% \boldmath$6.5<\qsq<100\,\gevt$}
\end{picture}
\caption{Differential cross sections $1/\sigma\,d\sigma / d\cos\theta^\ast$ in 
  $ep \rightarrow e\: J/\psi\: X$ in the kinematic region $50<W<225\,\mbox{~GeV}$, 
  $\ptstw>1\,\gevt$  and $ 0.3<z < 0.9$ normalised for $|\cos\theta^\ast|<0.9$.
  a)~$2 < Q^2 < 100\mbox{~GeV}^2$,
  b)~$2 < Q^2 < 6.5 \mbox{~GeV}^2$, c) $6.5 < Q^2 < 100 \mbox{~GeV}^2$.
  The inner error bars indicate the statistical  uncertainty, while
  the outer error bars include the statistical and  systematic uncertainties added
  in quadrature. The shaded regions show the result of fits with the form $\sim 1+\alpha
  \cos^2\theta^\ast$ and correspond to a variation  of the 
  fit parameter $\alpha$ by $\pm1$ standard deviation. The dashed lines are the 
  result of a prediction
  using the $k_t$ factorisation approach \cite{baranov}.}
 \label{thetastar}
\end{figure}

Measuring the polarisation of the produced $J/\psi$ meson has 
been proposed as a means of distinguishing the various CO and CS contributions to 
$J/\psi$ production~\cite{Kraemer,Fle972}.
The  polar ($\theta^\ast$) decay angular distributions are measured in the rest 
frame of the \jpsi\ using the \jpsi\ direction in the \gstp\ system as reference
axis (helicity frame). They are shown in Fig.~\ref{thetastar} (and listed in 
Table~\ref{costh_tab}) for the whole 
\qsq\ range and  separately for regions of low and high \qsq. 
The $\cos\theta^\ast$ distribution is expected to have the form 
\begin{equation}
  \frac{{\rm d}\,\sigma}{{\rm d}\,\cos\theta^{\ast}}\propto 1 + \alpha\, \cos^2\theta^{\ast}.\label{da}
\end{equation}

A value of $|\alpha|\lsim 0.5$  is 
expected, where $\alpha$  can be negative, zero or positive depending 
on which intermediate $c\bar{c}$ state dominates the production \cite{Fle972}.
Fitting the data distributions with a function of the form (\ref{da})
yields a value of  
%$\alpha=-0.08^{+0.29}_{-0.25}$ 
$\alpha=-0.1^{+0.4}_{-0.3}$ 
in the whole \qsq\ range (Fig.~\ref{thetastar}a). 
For $2<\qsq<6.5\,\gevt$ (Fig.~\ref{thetastar}b), 
%$\alpha=-0.41^{+0.32}_{-0.27}$ 
$\alpha=-0.4^{+0.5}_{-0.4}$ 
is found and 
for $6.5<\qsq<100\,\gevt$ (Fig.~\ref{thetastar}c) 
%$\alpha=0.67^{+0.65}_{-0.50}$
$\alpha=0.7^{+0.9}_{-0.6}$. The total
experimental errors were used in the fits.
%Note that only statistical errors are used in the fits. 
Although the central values  
suggest a change from a negative to a positive value of $\alpha$ 
as \qsq\ increases, this tendency is not significant.
Predictions using the $k_t$ factorisation approach \cite{baranov}, shown
in Fig.~\ref{thetastar}, 
are compatible with the measurements.
%-------------------------------------------------------------------------------

\section{Summary and Conclusions}
A new analysis of inelastic electroproduction of \jpsi\ mesons has
been presented. Due to the increased statistics the kinematic range has
been extended to $50<W<225\,\gev$ and 
reaches average values of \qsq\ larger than the squared
mass of the \jpsi\ meson.
The cross sections are measured in the range $0.3<z<0.9$ and
\ptst$>1\,\gevt$ where direct inelastic \jpsi\ production dominates.
%The remaining amount of background from diffractive \psits\ events
%and b decays is estimated to be $\sim 9- 17\%$.\\
% and can at present not be subtracted reliably.\\ 
Differential cross sections at average values  
$\langle Q^2\rangle = 10.6\; \mbox{and}\;30.9\mbox{~GeV}^2$ are
presented as functions of 
$Q^2$, $W$, $z$, $p_{t,\psi}^2$, $p_{t,\psi}^{*2}$, $Y$ and $Y^*$. 
Recent theoretical calculations by Kniehl and Zwirner \cite{kniehl} in 
the framework of the non--relativistic QCD (NRQCD)
approach including \coloct\ and \colsing\
contributions (``$2\ra 2$'' diagrams) are compared with the data.
At both average \qsq\ values, reasonable agreement is observed in the shapes of most
distributions except that of $z$, which is described much better by
the \colsing\ contribution alone 
(in a recent resummation of soft
            gluon processes a similar observation in the
            photoproduction limit could be explained through a
            damping of the rapid rise of the colour octet contributions
            towards high $z$ values).
%A similar observation in 
%the photoproduction limit was explained by resummation 
%of soft gluon processes, which damped the rapid rise of the 
%\coloct\ contributions towards high $z$ values.
%Thus this discrepancy is maybe not a serious problem in the
%present kinematic range either.  
The absolute value of the full NRQCD cross section is a factor $\sim 2$
above the data at low \qsq\ but approaches the data at higher \qsq\
to within 15\% which is well within experimental and theoretical uncertainties.
The \colsing\ contribution alone is  roughly a factor $2.7$ lower than the data.
The differential cross sections in \ptt\ and \ptst\ are
described better when  CO contributions are included. In the 
photoproduction
limit a successful description of the \ptt\ spectrum has been 
achieved within the \csm\ by including NLO corrections. These corrections are, 
however, not yet 
available for the electroproduction case under consideration here.
The dependence of the $\gamma^{\ast} p$ cross section on $W$ is the same, within errors, as in the photoproduction case.

Further distributions are studied in an attempt to assess the relative importance of 
the different CO and CS terms. Since their contributions are expected to vary with $z$, differential 
cross sections \dsdq\ and \dsdpts\ are measured in intervals of $z$. 
The shapes of the \ptst\ and \qsq\ spectra are found to be similar 
to those over the whole $z$ range although there is an indication of a steeper 
\qsq\ dependence at high $z$.
A fit to the distribution of the polar decay angle in the helicity frame
covering the whole \qsq\ range yields 
%$\alpha=-0.08^{+0.29}_{-0.25}$
$\alpha=-0.1^{+0.4}_{-0.3}$ for a parametrisation $1 + \alpha\,\cos^2\theta^{\ast}$.

Altogether the measurements presented here provide significant new information 
which will aid the  
further development of a quantitative understanding of \jpsi\ meson production within 
pQCD.
%%%%%%%%%%%%%%%%%%%%%%%%%%%%%%%%%%%%%%%%%%%%%%%%%%%%%%%%%%%%
\section*{Acknowledgements}

We are grateful to the HERA machine group whose outstanding
efforts have made and continue to make this experiment possible. 
We thank
the engineers and technicians for their work in constructing and now
maintaining the H1 detector, our funding agencies for 
financial support, the
DESY technical staff for continual assistance, 
and the DESY directorate for the
hospitality which they extend to the non DESY 
members of the collaboration. We want to thank B.A. Kniehl, 
L. Zwirner and S.P. Baranov for many discussions and for making their theoretical 
predictions available to us.
%-------------------------------------------------------------------------------
%\clearpage
%\include{biblio}

%\clearpage

\clearpage
%\include{lq_tabelfinal}
%\include{hq_tabelfinal}
%\include{2d_tabelfinal}

%\begin{landscape}
\setlength{\unitlength}{1cm}
\begin{table}[htbp]
 \begin{center}
% \begin{tabular}{|p{2.2cm}|c|l|}
 \begin{tabular}{|c|c|l|}
 \hline
 \multicolumn{3}{|c|}{\boldmath$e + p \rightarrow e + J/\psi + X$}\\ \hline \hline
\rule{0cm}{.5cm} {\bf\boldmath \qsq [GeV$^2$]}
 &{\bf$\mathbf{\langle \qsq \rangle}$\,[GeV$^2$]}
 &{\bf\boldmath $d\sigma/d\qsq$\boldmath[pb/GeV$^2$]}
 \\[0.1cm] \hline
$2.0$ -- $ 3.6$&$  2.8$&$ 48.9 \pm  5.3 \pm  7.8$\\ \hline
$ 3.6$ -- $ 6.5$&$  4.9$&$ 22.6 \pm  2.7 \pm  3.6$\\ \hline
$ 6.5$ -- $ 12.$&$  8.8$&$ 9.12 \pm 1.22 \pm 1.46$\\ \hline
$ 12.$ -- $ 20.$&$ 15.2$&$ 3.26 \pm 0.64 \pm 0.52$\\ \hline
$ 20.$ -- $ 40.$&$ 27.4$&$ 0.91 \pm 0.21 \pm 0.15$\\ \hline
$ 40.$ -- $100.$&$ 60.1$&$ 0.11 \pm 0.043 \pm 0.018$\\ \hline
 \hline
\rule{0cm}{.5cm} {\boldmath$z$} &{\boldmath$\langle z \rangle$}\ &{\bf\boldmath$d\sigma/dz$[pb]}
 \\[0.1cm] \hline
$0.30$ -- $0.45$&$0.375$&$ 306. \pm   56. \pm   64.$\\ \hline
$0.45$ -- $0.60$&$0.525$&$ 365. \pm   44. \pm   58.$\\ \hline
$0.60$ -- $0.75$&$0.675$&$ 392. \pm   41. \pm   63.$\\ \hline
$0.75$ -- $0.90$&$0.825$&$ 562. \pm   56. \pm   84.$\\ \hline
 \hline
\rule{0cm}{.5cm}{\bf\boldmath$W$[GeV]} &{\bf\boldmath$\langle W \rangle$\,[GeV]}&
{\bf\boldmath$d\sigma/dW$[pb/GeV]} \\[0.1cm] \hline
$ 50$ -- $ 70$&$ 61.9$&$ 2.28 \pm 0.49 \pm 0.36$\\ \hline
$ 70$ -- $ 95$&$ 83.7$&$ 1.70 \pm 0.23 \pm 0.27$\\ \hline
$ 95$ -- $120$&$106.9$&$ 1.71 \pm 0.20 \pm 0.27$\\ \hline
$120$ -- $145$&$132.7$&$ 1.34 \pm 0.18 \pm 0.22$\\ \hline
$145$ -- $170$&$156.8$&$ 1.21 \pm 0.20 \pm 0.19$\\ \hline
$170$ -- $195$&$181.2$&$ 1.04 \pm 0.23 \pm 0.17$\\ \hline
$195$ -- $225$&$207.4$&$ 1.01 \pm 0.26 \pm 0.16$\\ \hline
 \hline
\rule{0cm}{.5cm}{\bf\boldmath$p_{t,\psi}^{2}$[GeV$^2$]}&
{\bf\boldmath$\langle p_{t,\psi}^{2} \rangle$\,[GeV$^2$]} & 
{\bf\boldmath$d\sigma/dp_{t,\psi}^2$}{\bf\boldmath[pb/GeV$^2$]}
 \\[0.1cm] \hline
$ 0.0$ -- $ 1.0$&$  0.5$&$ 32.2 \pm  5.8 \pm  5.2$\\ \hline
$ 1.0$ -- $ 2.2$&$  1.5$&$ 25.0 \pm  4.6 \pm  4.0$\\ \hline
$ 2.2$ -- $ 3.7$&$  2.9$&$ 24.6 \pm  4.1 \pm  3.9$\\ \hline
$ 3.7$ -- $ 6.4$&$  5.1$&$ 14.7 \pm  2.4 \pm  2.4$\\ \hline
$ 6.4$ -- $ 9.6$&$  8.0$&$ 8.42 \pm 1.69 \pm 1.35$\\ \hline
$ 9.6$ -- $13.5$&$ 11.2$&$ 7.38 \pm 1.22 \pm 1.18$\\ \hline
$13.5$ -- $ 20.$&$ 16.3$&$ 3.58 \pm 0.63 \pm 0.57$\\ \hline
$ 20.$ -- $ 40.$&$ 26.2$&$ 0.98 \pm 0.17 \pm 0.16$\\ \hline
$ 40.$ -- $100.$&$ 55.5$&$0.088 \pm 0.028 \pm 0.014$\\ \hline
 \end{tabular}
 \end{center}
 \caption{Differential cross sections with statistical and systematic
 errors in the range 
$\mbox{$2<\qsq<100$\,GeV$^2$}$, $50<W<225\,\mbox{~GeV}$,
 $0.3<z<0.9$ and $p_{t,\psi}^{\ast 2}>1\,\rm GeV^2$.
 \label{lq_tab1}}
 \end{table}

 \begin{table}
 \begin{center}
%\begin{tabular}{|p{2.2cm}|c|l|}
 \begin{tabular}{|c|c|l|}
 \hline
 \multicolumn{3}{|c|}{\boldmath$e + p \rightarrow e + J/\psi + X$}\\ \hline \hline
 \rule{0cm}{.5cm} {\bf\boldmath$Y^\ast$}& {\bf\boldmath $\langle Y^\ast \rangle$}
 & {\bf\boldmath$d\sigma/dY^\ast$ [pb]} \\[0.1cm] \hline
$2.00$ -- $2.40$&$2.26$&$ 100. \pm   18. \pm   16.$\\ \hline
$2.40$ -- $2.65$&$2.54$&$ 137. \pm   24. \pm   22.$\\ \hline
$2.65$ -- $2.90$&$2.78$&$ 167. \pm   23. \pm   27.$\\ \hline
$2.90$ -- $3.15$&$3.02$&$ 185. \pm   21. \pm   30.$\\ \hline
$3.15$ -- $3.40$&$3.27$&$ 127. \pm   18. \pm   20.$\\ \hline
$3.40$ -- $4.00$&$3.57$&$ 70.5 \pm  11.1 \pm  11.3$\\ \hline
 \hline
 \rule{0cm}{.5cm} {\bf\boldmath$p_{t,\psi}^{\ast 2}$  [GeV$^2$]}
 & {\bf\boldmath$\langle p_{t,\psi}^{\ast 2} \rangle$\,[GeV$^2$]}
 &{\bf\boldmath$d\sigma/dp_{t,\psi}^{\ast 2}$[pb/GeV$^2$]} \\[0.1cm] \hline
$ 1.0$ -- $ 2.2$&$  1.6$&$ 61.2 \pm  7.2 \pm  9.8$\\ \hline
$ 2.2$ -- $ 3.7$&$  2.8$&$ 36.6 \pm  4.9 \pm  5.5$\\ \hline
$ 3.7$ -- $ 6.4$&$  4.9$&$ 18.9 \pm  2.6 \pm  2.8$\\ \hline
$ 6.4$ -- $ 9.6$&$  7.8$&$ 7.48 \pm 1.43 \pm 1.12$\\ \hline
$ 9.6$ -- $13.5$&$ 11.0$&$ 3.20 \pm 0.82 \pm 0.48$\\ \hline
$13.5$ -- $ 20.$&$ 16.4$&$ 1.92 \pm 0.44 \pm 0.29$\\ \hline
$ 20.$ -- $ 40.$&$ 26.6$&$ 0.57 \pm 0.15 \pm 0.085$\\ \hline
 \hline
  \rule{0cm}{.5cm} {\boldmath$Y_{lab}$}
 &{\boldmath$\langle Y_{lab} \rangle$}&{\boldmath$d\sigma/dY_{lab}$}{\bf[pb]}
 \\[0.1cm] \hline
$-1.5$ -- $-0.8$&$-1.02$&$ 70.8 \pm  10.1 \pm  11.3$\\ \hline
$-0.8$ -- $-0.5$&$-0.66$&$ 98.7 \pm  15.3 \pm  15.8$\\ \hline
$-0.5$ -- $-0.2$&$-0.34$&$ 78.0 \pm  13.4 \pm  12.5$\\ \hline
$-0.2$ -- +$ 0.1$&$-0.06$&$ 92.5 \pm  14.0 \pm  14.8$\\ \hline
$ 0.1$ -- $ 0.4$&$ 0.24$&$ 83.5 \pm  13.8 \pm  13.4$\\ \hline
$ 0.4$ -- $ 0.7$&$ 0.54$&$ 66.2 \pm  14.0 \pm  10.6$\\ \hline
$ 0.7$ -- $ 1.0$&$ 0.84$&$ 94.8 \pm  17.1 \pm  15.2$\\ \hline
$ 1.0$ -- $ 1.5$&$ 1.16$&$ 45.7 \pm  14.0 \pm   7.3$\\ \hline
 \end{tabular}
 \end{center}
 \caption{Differential cross sections with statistical and systematic errors in the range 
  $\mbox{$2<\qsq<100$\,GeV$^2$}$, $50<W<225\,\mbox{~GeV}$,
 $0.3<z<0.9$ and $p_{t,\psi}^{\ast 2}>1\,\rm GeV^2$.
 \label{lq_tab2}}
 \end{table}
 
 \begin{table}[htbp]
 \begin{center}
% \begin{tabular}{|p{2.5cm}|c|l|}
 \begin{tabular}{|c|c|l|}
 \hline
 \multicolumn{3}{|c|}{\boldmath$e + p \rightarrow e + J/\psi + X$}\\
 \hline \hline
\rule{0cm}{.5cm} {\bf\boldmath \qsq [GeV$^2$]}
 &{\bf$\mathbf{\langle \qsq \rangle}$\,[GeV$^2$]}
 &{\bf\boldmath $d\sigma/d\qsq$\boldmath[pb/GeV$^2$]}
 \\[0.1cm] \hline
$ 12$ -- $ 20$&$ 15.3$&$ 1.87 \pm 0.44 \pm 0.30$\\ \hline
$ 20$ -- $ 40$&$ 28.5$&$ 0.60 \pm 0.16 \pm 0.096$\\ \hline
$ 40$ -- $100$&$ 59.8$&$0.092 \pm 0.036 \pm 0.015$\\ \hline
 \hline 
\rule{0cm}{.5cm} {\boldmath$z$} &{\boldmath$\langle z \rangle$}\ &{\bf\boldmath$d\sigma/dz$[pb]}
 \\[0.1cm] \hline
$0.30$ -- $0.60$&$0.450$&$ 48.3 \pm  13.6 \pm   8.7$\\ \hline
$0.60$ -- $0.75$&$0.675$&$ 46.6 \pm  14.9 \pm   7.5$\\ \hline
$0.75$ -- $0.90$&$0.825$&$ 52.7 \pm  15.2 \pm   7.9$\\ \hline
 \hline
\rule{0cm}{.5cm}{\bf\boldmath$W$[GeV]} &{\bf\boldmath$\langle W \rangle$\,[GeV]}&
{\bf\boldmath$d\sigma/dW$[pb/GeV]} \\[0.1cm] \hline
$ 50$ -- $ 85$&$ 78.5$&$ 0.47 \pm  0.17 \pm 0.075$\\ \hline
$ 85$ -- $110$&$ 98.9$&$ 0.26 \pm 0.086 \pm 0.041$\\ \hline
$110$ -- $135$&$125.9$&$ 0.20 \pm 0.069 \pm 0.032$\\ \hline
$135$ -- $225$&$169.7$&$ 0.12 \pm 0.033 \pm 0.019$\\ \hline
 \hline
 \hline
\rule{0cm}{.5cm}{\bf\boldmath$p_{t,\psi}^{2}$[GeV$^2$]}&
{\bf\boldmath$\langle p_{t,\psi}^{2} \rangle$\,[GeV$^2$]} & 
{\bf\boldmath$d\sigma/dp_{t,\psi}^2$}{\bf\boldmath[pb/GeV$^2$]}
 \\[0.1cm] \hline
$ 6.4$ -- $13.5$&$  9.6$&$ 2.13 \pm 0.59 \pm 0.34$\\ \hline
$13.5$ -- $ 40.$&$ 21.4$&$ 0.52 \pm 0.12 \pm 0.084$\\ \hline
$ 40.$ -- $100.$&$ 56.0$&$0.052 \pm 0.022 \pm 0.008$\\ \hline
 \hline
 \rule{0cm}{.5cm} {\bf\boldmath$Y^\ast$}& {\bf\boldmath $\langle Y^\ast \rangle$}
 & {\bf\boldmath$d\sigma/dY^\ast$ [pb]}
 \\[0.1cm] \hline
$2.40$ -- $2.65$&$2.56$&$ 33.5 \pm  10.1 \pm   5.4$\\ \hline
$2.65$ -- $2.90$&$2.80$&$ 19.7 \pm   7.9 \pm   3.2$\\ \hline
$2.90$ -- $3.15$&$3.03$&$ 34.3 \pm   8.7 \pm   5.5$\\ \hline
$3.15$ -- $3.40$&$3.31$&$ 11.1 \pm   6.0 \pm   1.8$\\ \hline
 \hline
 \rule{0cm}{.5cm} {\bf\boldmath$p_{t,\psi}^{\ast 2}$  [GeV$^2$]}
 & {\bf\boldmath$\langle p_{t,\psi}^{\ast 2} \rangle$\,[GeV$^2$]}
 &{\bf\boldmath$d\sigma/dp_{t,\psi}^{\ast 2}$[pb/GeV$^2$]}
 \\[0.1cm] \hline
$ 1.0$ -- $ 2.2$&$  1.5$&$ 5.48 \pm 2.10 \pm 0.88$\\ \hline
$ 2.2$ -- $ 6.4$&$  4.0$&$ 3.58 \pm 0.88 \pm 0.54$\\ \hline
$ 6.4$ -- $ 20.$&$ 10.2$&$ 0.54 \pm  0.18 \pm 0.081$\\ \hline
$ 20.$ -- $ 40.$&$ 28.5$&$ 0.12 \pm 0.074 \pm 0.018$\\ \hline
 \end{tabular}
 \end{center}
 \caption{Differential cross sections with statistical and systematic 
errors in the range
  $\mbox{$12<\qsq<100$\,GeV$^2$}$, $50<W<225\,\mbox{~GeV}$,
 $p_{t,\psi}^{2}>6.4\rm GeV^2$,
 $0.3<z<0.9$ and $p_{t,\psi}^{\ast 2}>1\,\rm GeV^2$.
 \label{hq_tab}}
 \end{table}

 \begin{table}[htbp]
 \begin{center}
 \begin{tabular}{|p{2.5cm}|c|l|}
 \hline
 \multicolumn{3}{|c|}{\boldmath$e + p \rightarrow e + J/\psi + X$}\\ \hline \hline
 \rule{0cm}{.5cm} {\bf\boldmath$p_{t,\psi}^{\ast 2}$ [GeV$^2$]}
 & {\bf\boldmath$\langle p_{t,\psi}^{\ast 2} \rangle$\,[GeV$^2$]}
 &{\bf\boldmath$d\sigma/dp_{t,\psi}^{\ast 2} dz$[pb/GeV$^2$]}
  \\[0.1cm] \hline
\multicolumn{3}{|c|}{$0.30<z<0.60, \ \langle z\rangle=0.45 $}
 \\ \hline
$ 1.0$ -- $ 2.2$&$  1.5$&$ 81.8\pm  16.6 \pm  17.2$\\ \hline
$ 2.2$ -- $ 6.4$&$  3.8$&$ 36.3 \pm   5.7 \pm   7.6$\\ \hline
$ 6.4$ -- $13.5$&$  8.8$&$ 7.76 \pm  1.89 \pm  1.63$\\ \hline
$13.5$ -- $ 40.$&$ 21.2$&$ 1.18 \pm  0.33 \pm  0.25$\\ \hline
\multicolumn{3}{|c|}{$0.60<z<0.75, \ \langle z\rangle=0.675 $}
 \\ \hline
$ 1.0$ -- $ 2.2$&$   1.6$&$ 116.7 \pm  20.1 \pm  24.5$\\ \hline
$ 2.2$ -- $ 6.4$&$  3.6$&$  37.9 \pm   6.4 \pm   8.0$\\ \hline
$ 6.4$ -- $13.5$&$   8.6$&$ 5.84 \pm  2.07 \pm  1.23$\\ \hline
$13.5$ -- $ 40.$&$ 19.7$&$ 1.48 \pm  0.48 \pm  0.31$\\ \hline
\multicolumn{3}{|c|}{$0.75<z<0.90, \ \langle z\rangle=0.825 $}
 \\ \hline
$ 1.0$ -- $ 2.2$&$  1.6$&$ 113.6 \pm  24.4 \pm  23.9$\\ \hline
$ 2.2$ -- $ 6.4$&$  3.8$&$  57.0 \pm   8.8 \pm  12.0$\\ \hline
$ 6.4$ -- $13.5$&$  9.3$&$  10.8 \pm   3.0 \pm   2.3$\\ \hline
$13.5$ -- $ 40.$&$ 22.6$&$ 2.33 \pm  0.65 \pm  0.49$\\ \hline
 \hline
 \rule{0cm}{.5cm} {\bf\boldmath \qsq [GeV$^2$]}
 &{\bf$\mathbf{\langle \qsq \rangle}$\,[GeV$^2$]}
 &{\bf\boldmath $d\sigma/d\qsq dz$\boldmath[pb/GeV$^2$]}
 \\[0.1cm] \hline
\multicolumn{3}{|c|}{$0.30<z<0.60, \ \langle z\rangle=0.45 $}
 \\ \hline
$ 2.0$ -- $ 6.5$&$  3.8$&$ 38.9 \pm  5.4 \pm  8.2$\\ \hline
$ 6.5$ -- $ 20.$&$ 11.1$&$ 9.47 \pm 1.57 \pm 1.99$\\ \hline
\multicolumn{3}{|c|}{$0.60<z<0.75, \ \langle z\rangle=0.675 $}
 \\ \hline
$ 2.0$ -- $ 6.5$&$  3.5$&$ 54.8 \pm   7.3 \pm  11.5$\\ \hline
$ 6.5$ -- $ 20.$&$ 11.3$&$ 7.61 \pm  1.58 \pm  1.60$\\ \hline
$ 20.$ -- $100.$&$ 39.1$&$ 0.38 \pm 0.155 \pm 0.079$\\ \hline
\multicolumn{3}{|c|}{$0.75<z<0.90, \ \langle z\rangle=0.825 $}
 \\ \hline
$ 2.0$ -- $ 6.5$&$  3.7$&$ 82.6 \pm  10.4 \pm  17.3$\\ \hline
$ 6.5$ -- $ 20.$&$ 11.1$&$ 9.25 \pm  2.09 \pm  1.94$\\ \hline
$ 20.$ -- $100.$&$ 36.9$&$ 0.52 \pm  0.17 \pm  0.11$\\ \hline
 \end{tabular}
 \end{center}
 \caption{Double differential cross sections with statistical and
 systematic errors in the range 
 $2<\qsq<100$\,GeV$^2$, $50<W<225\,\mbox{~GeV}$,
 $0.3<z<0.9$ and $p_{t,\psi}^{\ast 2}>1\,\rm GeV^2$.
 \label{2d_tab}}
 \end{table}

\begin{table}[h]
\begin{center}
% \begin{tabular}{|p{2.cm}|c|l|}
 \begin{tabular}{|c|c|l|}
 \hline
 \multicolumn{3}{|c|}{\boldmath$\gamma^\ast + p \rightarrow J/\psi + X$}\\ \hline \hline
\rule{0cm}{.5cm}{\bf\boldmath$W$[GeV]} &{\bf\boldmath$\langle W \rangle$\,[GeV]}&
\multicolumn{1}{|c|}{\bf\boldmath$\sigma_{\gamma^\ast p}$[pb]} \\[0.1cm] \hline
$ 50$ -- $ 70$&$ 59.3$&$ 7.78 \pm 1.68 \pm 1.24$\\ \hline
$ 70$ -- $ 95$&$ 81.7$&$ 8.25 \pm 1.12 \pm 1.32$\\ \hline
$ 95$ -- $120$&$106.9$&$ 11.3 \pm  1.3 \pm  1.8$\\ \hline
$120$ -- $145$&$131.9$&$ 11.6 \pm  1.6 \pm  1.9$\\ \hline
$145$ -- $170$&$157.0$&$ 13.4 \pm  2.3 \pm  2.1$\\ \hline
$170$ -- $195$&$182.0$&$ 14.5 \pm  3.2 \pm  2.3$\\ \hline
$195$ -- $225$&$209.3$&$ 17.8 \pm  4.5 \pm  2.9$\\ \hline
 \end{tabular}
 \end{center}
 \caption{Cross sections with statistical and systematic errors for
 $\gamma^\ast + p \rightarrow J/\psi + X$ at 
$\mbox{$\langle Q^2\rangle\simeq 10.6\mbox{~GeV}^2$}$
 in the range $0.3<z<0.9$ and $p_{t,\psi}^{\ast 2}>1\,\rm GeV^2$.
 \label{gp_tab1}}
% \end{table}

%\begin{table}[h]
 \begin{center}
% \begin{tabular}{|p{2.5cm}|c|c|c|c|}
 \begin{tabular}{|c|c|c|c|c|}
 \hline
 \multicolumn{3}{|c|}{\boldmath$e + p \rightarrow e + J/\psi + X$}\\ \hline \hline
 {\boldmath$\cos(\theta^\ast)$}
 &{\boldmath$\langle \cos(\theta^\ast) \rangle$}
 &{\bf\boldmath$d\sigma/d\cos(\theta^\ast)$[pb]} \\[0.1cm] \hline
 \multicolumn{3}{|c|}{$2<\qsq<100\, \rm GeV^2$}\\ \hline
$-0.90$ -- $-0.55$&$-0.70$&$ 111. \pm   20. \pm   18.$\\ \hline
$-0.55$ -- $-0.33$&$-0.43$&$ 141. \pm   21. \pm   22.$\\ \hline
$-0.33$ -- $-0.11$&$-0.21$&$ 108. \pm   18. \pm   17.$\\ \hline
$-0.11$ -- +$0.11$&$-0.01$&$ 112. \pm   18. \pm   18.$\\ \hline
$0.11$ -- $0.33$&$0.24$&$ 155. \pm   20. \pm   25.$\\ \hline
$0.33$ -- $0.55$&$0.45$&$ 125. \pm   20. \pm   20.$\\ \hline
$0.55$ -- $0.90$&$0.70$&$ 122. \pm   22. \pm   20.$\\ \hline
 \hline
 \multicolumn{3}{|c|}{$2<\qsq<6.5\,\rm GeV^2$}\\ \hline
$-0.90$ -- $-0.45$&$-0.64$&$ 55.5 \pm  12.2 \pm   8.9$\\ \hline
$-0.45$ -- $-0.20$&$-0.35$&$ 56.7 \pm  13.0 \pm   9.1$\\ \hline
$-0.20$ -- +$0.20$&$-0.01$&$ 74.8 \pm  10.2 \pm  12.0$\\ \hline
$0.20$ -- $0.45$&$0.32$&$116.0 \pm  16.1 \pm  18.6$\\ \hline
$0.45$ -- $0.90$&$0.65$&$ 70.9 \pm  13.7 \pm  11.3$\\ \hline
 \hline
 \multicolumn{3}{|c|}{$6.5<\qsq<100\, \rm GeV^2$}\\ \hline
$-0.90$ -- $-0.45$&$-0.66$&$ 52.4 \pm  11.0 \pm   8.4$\\ \hline
$-0.45$ -- $-0.20$&$-0.33$&$ 68.6 \pm  12.8 \pm  11.0$\\ \hline
$-0.20$ -- +$0.20$&$-0.02$&$ 37.8 \pm   8.3 \pm   6.0$\\ \hline
$0.20$ -- $0.45$&$0.30$&$ 47.7 \pm  11.5 \pm   7.6$\\ \hline
$0.45$ -- $0.90$&$0.64$&$ 59.5 \pm  12.4 \pm   9.5$\\ \hline
 \end{tabular}
 \end{center}
 \caption{Differential cross sections with statistical and systematic
 errors in the range 
 $\mbox{$2<\qsq<100$\,GeV$^2$}$, $50<W<225\,\mbox{~GeV}$,
 $0.3<z<0.9$ and $p_{t,\psi}^{\ast 2}>1\,\rm GeV^2$.
 \label{costh_tab}}
 \end{table}

\end{document}